\documentclass[aps,prd,preprint,groupedaddress]{revtex4-2}
\usepackage{epsfig}
\usepackage{array,tabularx,epsfig,mathrsfs,graphicx,rotating}
\usepackage{ifthen}
\usepackage{ragged2e}
\PassOptionsToPackage{hyphens}{url}
\usepackage[hyphens]{url}
\usepackage{hyperref}
\usepackage{listings}
\usepackage{epstopdf}
\usepackage{color}
\usepackage{float}
\usepackage{subfig}
\usepackage{makecell}
\usepackage{xspace}

\usepackage{ulem}
\usepackage{amssymb}
\usepackage{booktabs}
\usepackage{graphicx}% Include figure files
\usepackage[export]{adjustbox} % provides valign=m
\usepackage{physics}
\usepackage{dcolumn}% Align table columns on decimal point
\usepackage{bm}% bold math
\usepackage{xcolor}
\usepackage{comment}

\usepackage{amsmath}
\usepackage{amsfonts}
\usepackage{dsfont}
\usepackage{tikz}

\newcommand{\Cov}{\textnormal{Cov}}
\newcommand{\Var}{\textnormal{Var}}
\newcommand{\sigprime}{\sigma_{\textnormal{reweight}}}
\newcommand{\poisson}{\mathds{1}}
\hypersetup{
  colorlinks=true,
  linkcolor=blue,
  citecolor=blue,
  urlcolor=blue
}

\newcommand{\pt}{\ensuremath{p_{\textnormal{T}}}}
\newcommand{\ptV}{\ensuremath{\pt(\textnormal{V})}}

\newcommand{\NNPDF}{\textsc{NNPDF-NLO}\xspace} % from ptdr-definitions

\newcommand{\vjets}{\ensuremath{V+\mathrm{jets}}\xspace}

\newcommand{\dnnsvb} {\ensuremath{\mathrm{DNN}_{\mathrm{SvB}}}\xspace}
\newcommand{\dnnre} {\ensuremath{\mathrm{DNN}_{\mathrm{re}}}\xspace}

\begin{document}

%\begin{frontmatter}

\title{Improving statistical precision in Monte Carlo samples with negative weights via reweighting and uncertainty quantification}
\author{C.~Palmer}
\author{B.~Kronheim}

% I've added not in alpahetic order, not just because I want my name the first (I do not care), but
% to make as less change as possible since this paper was sitting in the public domain 
% with the current old list  of the authors for a long time,
% so this change will be less visible from outside.

%\address[a]{Department of  Physics, U. Maryland,
%             College Park,
%             Maryland,
%             MD 20742,
%             USA}

\affiliation{Department of Physics, University of Maryland, College Park, Maryland, 20742, USA}

%\emailAdd{chekanov@anl.gov}
%\note{Preprint ANL-HEP-186226}

\date{\today}

\begin{abstract}
High statistical precision is critical for Monte Carlo (MC) samples in high energy physics and is degraded by negatively weighted events. This paper investigates a procedure to learn the relationship between the negative and positive weight distributions of any sample, allowing the reduction of statistical uncertainty by reweighting kinematically equivalent events with the same sign. A robust uncertainty quantification method is required for the practical application of such  method. Two methods for the estimation of the reweighting uncertainty are developed: one at the event and another at the final observable level. The latter method is strongly favored. The gains in statistical precision are then quantified. The method is demonstrated on Sherpa vector boson plus jets samples when using all generated events and when restricted to the signal region of a mock analysis. It is demonstrated to significantly reduce stochastic behavior in sparse MC samples while decreasing the overall uncertainty with a sufficiently well-known reweighting function.
\end{abstract}

\maketitle

\noindent Submitted to Phys. Rev. D:  November 4th, 2025 \\ 
Accepted by Phys. Rev. D:  December 12th. 2025

\clearpage

%\arxivnumber{XXXXX.XXXXX} % Only if you have one

%% Keywords
%\begin{keyword}
%Simulations \sep Machine Learning \sep Collider Physics \sep 
%% keywords here, in the form: keyword \sep keyword
%% PACS codes here, in the form: \PACS code \sep code
%% MSC codes here, in the form: \MSC code \sep code
%% or \MSC[2008] code \sep code (2000 is the default)
%\end{keyword}

%\end{frontmatter}

\section{\label{sec:intro}Introduction}

Analyses in high energy physics (HEP) depend on the accurate modeling of signal and background processes. The modeling is generally done in whole or in part with Monte Carlo (MC) simulations, which produce samples of synthetic events for analysis. Matrix element (ME) generators such as Madgraph5 \cite{Alwall:2011uj}, aMC@NLO \cite{Alwall:2014hca}, Powheg \cite{Frixione:2007vw}, Powheg-Box \cite{Alioli:2010xd}, and Sherpa \cite{Bothmann:2024eg} sample from the allowed kinematics of a hard scattering process such as $u\bar{u}\to Zb\bar{b}\to \mu\bar{\mu}b\bar{b}$. Parton shower (PS) generators such as Pythia8 \cite{Bierlich:2022pfr} and Herwig7 \cite{Bellm:2015jjp} receive the particles in events from ME generators and simulate soft quantum electrodynamic and chromodynamic emissions and splittings. Typically, the parton showers are paired with hadronizers to combine the showered quarks and gluons into mesons and baryons. 

Generators can now provide next-to-leading order (aMC@NLO, Powheg-Box, Sherpa) and even next-to-next-to-leading order (MINNLO$_{PS}$ \cite{Monni:2019whf}) accuracies. The higher accuracy of these predictions typically comes with non-negligible fractions of negative weights. These negative weights come from many sources, ranging from interference terms to the handling of soft and collinear emissions shared between the ME and PS levels. Regardless of the source, their impact is the same: larger MC statistical uncertainties in sensitive analysis regions and reduced sensitivity in physics searches and measurements. 

Large statistical fluctuations are expected in sparsely populated bins for all MC samples, but for the same size sample many more bins are affected and the fluctuations can be negative when there is a large fraction of negative weights. Significantly larger samples are needed to compensate for the negative weights and achieve the same statistical precision as samples with no (or very few) negatively weighted events, such as most leading-order (LO) samples. Ensuring similar statistical precision can easily inflate the required next-to-leading order (NLO) sample sizes to be many times that of the LO samples.

Previous works have attempted to resample already produced distributions \cite{Andersen:2020pr, Andersen:2022ue, Nachman:2020fff}, and others have used neural networks to map between physics distributions \cite{Andreassen:2019nnm}. A contemporary paper~\cite{Nachman:2025sp} proposed a negative refinement method to remove or reduce negative weights via reweighting. This paper's similar reweighting method was simultaneously developed in parallel. In addition, Ref.~\cite{Nachman:2025sp} is expanded upon with new contributions on the impacts of the reweighting on statistical uncertainty and the proper handling of uncertainties on the reweighting. This paper starts with a discussion on the experimental consequences of negative weights in Sec.~\ref{sec:consequence_of_negative} and their origin in Sec.~\ref{sec:origin_of_negative}. 
In Sec.~\ref{sec:reweighting_derivation} the reweighting function is derived and its application strategy is described in Sec.~\ref{sec:application_strategy}. 
In Sec.~\ref{sec:reweightnoUncert} the case of exact reweighting is discussed and it is applied using the double slit experiment in Sec.~\ref{sec:double_slit} to demonstrate the significant gains in statistical uncertainty when the reweighting is known exactly. 
In Sec.~\ref{sec:uncertainty_quantification} the situation where the reweighting function is less precisely known is addressed. This paper presents a method for estimating the reweighting function in Sec.~\ref{sec:model_choice} and proposes a method for estimating the associated uncertainties in Sec.~\ref{sec:uncertainty_properties} using an established ensemble method \cite{Efron:1992spr, Efron:1994ib}. A prescription for applying the uncertainties event-by-event is presented in Sec.~\ref{sec:event_uncertainties} and at the final observable level in Sec.~\ref{sec:histogram_uncertainties}.

The reweighting method is applied to a vector boson plus jets Sherpa sample \cite{ATLAS:electroweakBoson} released by ATLAS \cite{ATLAS:2008xda} via CERN Open Data. The sample has about $70\%$ positive weights when there are at least two partons and the scalar sum of the transverse momentum of the partons and vector boson is relatively large, the main region of interest for studies such as $VH\to Vb\bar{b}$. It is used as the background to an ATLAS vector boson plus Higgs boson sample \cite{ATLAS:higgs}. The closure of the reweighting, the sizes of the estimated uncertainties, and the expected sensitivity gains are then demonstrated using these samples in Sec.~\ref{sec:VH_example}. The samples are described in Sec.~\ref{sec:samples}, the details of the reweighting model are explained in Sec.~\ref{sec:model_training}, and the closure on training variables is shown in Sec.~\ref{sec:training_variable_closure}. The signal region definition is presented in Sec.~\ref{sec:signal_region_definition} and the result of the reweighting is included in Sec.~\ref{sec:PCA_systematics}. Finally, the conclusions are presented in Sec.~\ref{sec:conclusion}.

\subsection{\label{sec:consequence_of_negative}Consequences of negative weights}

The largest issue with samples containing negative weights is that they require substantially more generated events to obtain the same precision compared to samples with purely positively weighted events. Specifically, with a fraction of positive events $P_{+}$, the number of generated events must be scaled by
\begin{equation}
    f\left(P_{+}\right) = \frac{1}{\left(2P_{+}-1\right)^2},
\label{eqn:equivSamples}
\end{equation}
as shown in Refs.~\cite{Frederix2020,Danziger2021Reducing}. Compared to a purely positive sample, one with a positive fraction of 0.9 requires more than $50\%$ additional events to obtain the same statistical precision. With the increasing luminosity of the LHC, particularly with the upcoming upgrade of the LHC to the High Luminosity LHC, generating and storing the required number of samples becomes even more challenging. Samples are also often analyzed separately according to their flavor content. For example, a \vjets sample may be split into light flavored events, those with c jets, and those with b jets, further increasing the number of events necessary for smooth histograms.

\subsection{\label{sec:origin_of_negative}Origin of negative weights}

In order to implement a general method to mitigate negative weights, their source must be understood. Through understanding the source, the generator and parton shower variables responsible for the negative weights can be identified and used in the mitigation method. Using these variables is particularly advantageous for a reweighting-based method, as reweighting these variables ensures proper modeling of all derived variables, including all reconstruction-level variables. Using these variables allows the use of nearly all event attributes in a reweighted sample. The paper focuses on the sources of negative weights in Sherpa, as these are the samples used in this paper. For other generators (e.g., aMC@NLO), the exact details will be different, resulting in potentially distinct optimal variables. 

Negative weights occur in NLO calculations due to the computational infrared subtraction techniques that isolate divergences in real emissions. These techniques generally sum over spin and color related quantities in $m + 1$ partons to cancel the divergences via an equal but oppositely signed term in $m$ partons. While this technique leads to a finite result, the individual integrals can be negative in parts of the phase space~\cite{CATANI1997291}. 

Jets from the PS and ME level can share the same kinematic phase space and lead to double counting. Properly accounting for this can introduce additional negative event weights. In Sherpa, the real-emission phase space is divided into hard and soft regions, with the soft region constructed to fully cancel the subtraction term~\cite{Hoche2013}. As a result, the structure of the splitting function is too complex to use directly because it must be integrated and inverted. Sherpa instead generates splitting terms using an alternative function. The function is adjusted to the target with a weight, which is negative if the approximation is larger than the target in any region~\cite{Hoeche:2012yf}.

The most important variables are therefore those relating to the outgoing partons and the hard scatter itself. The exact variables used will be discussed in Sec.~\ref{sec:model_training}, but it is already evident that directly providing some information beyond the four vectors from the hard scatter will aid the method in different kinematic regions. In particular, information about the source of the partons is critical, as the origin of the partons can determine how they contribute to the sign fo the weight. This information will generally be calculated from the four vectors, but in a non-trivial fashion. Thus, directly providing it will ease the burden on the model learning it.

\section{\label{sec:reweighting_derivation}Reweighting Method Derivation}

The reweighting method applied in this paper is derived in full generality in the following text, and all assumptions are stated. While the method is applied to high energy physics MC event generators here, it is applicable wherever two independent probability distribution functions (PDFs) need to be combined. The uncertainty analysis which follows assumes the PDFs are produced through sampling.

%MC physics event generators predict final state kinematics and energy-momentum four-vectors of particles resulting from particle collisions at high energy accelerators from the beam type and four-momentum. The vector $\vec{x}$ is defined to have all relevant incoming and outgoing kinematic variables that determine the MC weight. 

Let the vector $\vec{x}$ have a generic set of input features from the MC generation space. Assume that the positively and negatively weighted events come from fixed probability distributions: PDF$_+\left(\vec{x}\right)$ and PDF$_-\left(\vec{x}\right)$. Assume that the full distribution has net integral 1 and can be written as a linear combination of these PDFs. Specifically,
\begin{equation}
    \textnormal{PDF}(\vec{x}) = a\textnormal{PDF}_+\left(\vec{x}\right) - b\textnormal{PDF}_-\left(\vec{x}\right)
\end{equation}
where $a, b \geq 0,$ $a>b,$ and $a - b = 1.$

The probability of obtaining a positively weighted event at any $\vec{x}$ can be expressed in terms of the already defined PDFs: 
\begin{equation}
P_+(\vec{x})=\frac{a\textnormal{PDF}_+(\vec{x})}{a\textnormal{PDF}_+(\vec{x})+b\textnormal{PDF}_-(\vec{x})}.
\end{equation}
The overall PDF can then be written as
\begin{equation}
\begin{aligned}
\textnormal{PDF}(\vec{x}) &= a\textnormal{PDF}_+(\vec{x}) - b\textnormal{PDF}_-(\vec{x})\\
&=a\textnormal{PDF}_+(\vec{x})\frac{a\textnormal{PDF}_+(\vec{x})+b\textnormal{PDF}_-(\vec{x})}{a\textnormal{PDF}_+(\vec{x})+b\textnormal{PDF}_-(\vec{x})} - b\textnormal{PDF}_-(\vec{x})\frac{a\textnormal{PDF}_+(\vec{x})+b\textnormal{PDF}_-(\vec{x})}{a\textnormal{PDF}_+(\vec{x})+b\textnormal{PDF}_-(\vec{x})}\\
&= (P_+(\vec{x}) - P_-(\vec{x}))(a\textnormal{PDF}_+(\vec{x})+b\textnormal{PDF}_-(\vec{x}))\\
&= \left(2P_+(\vec{x})-1\right)(a\textnormal{PDF}_+(\vec{x})+b\textnormal{PDF}_-(\vec{x}))
\end{aligned}.
\end{equation}

From the final equation, define
\begin{equation}
g(\vec{x})=2P_+(\vec{x})-1.
\end{equation}
Thus, the reweighted PDF is defined as
\begin{equation}
\textnormal{PDF}_\textnormal{reweight}(\vec{x})= g(\vec{x})(a\textnormal{PDF}_+(\vec{x}) + b\textnormal{PDF}_-(\vec{x})).
\end{equation}
This equation expresses the PDF in terms of the summed absolute value of the weights scaled by a fixed factor determined by the location in feature space. By construction it is equal to the original construction, so $\textnormal{PDF} = \textnormal{PDF}_\textnormal{reweight}$.

These formulas are valid for all points in feature space because no requirements are placed on $\vec{x}$ or $P_+(\vec{x})$. Specifically, $P_+(\vec{x})$ can have any value in $[0,1]$, and $\vec{x}$ may be any point in feature space.  Moreover, any feature $y$ that depends solely on $\vec{x}$ will be accurately reproduced by $\textnormal{PDF}_\textnormal{reweight}$. If the distribution of $y$ fails validation, then it must depend on additional variables not captured by $\vec{x}$.

There are also no requirements on how the MC samples from the PDFs are generated or how individual events are weighted. Indeed, the only requirement is that the distributions from which the positive and negative samples are generated are fixed. This requirement is trivially satisfied by any MC generator. For the upcoming uncertainty analysis, however, the independence of all generated events will be assumed.

There are several properties of $g(\vec{x})$ that are worth mentioning. 
First, $g(\vec{x}_i)$ can equal $g(\vec{x}_j)$ where $\vec{x}_i\neq\vec{x}_j$, as different regions can have the same degree of cancellation. This is expected for cases when the simulation covers multiple disjoint feature spaces. It is not assumed to be continuous or differentiable anywhere. Next, as PDFs are positive by definition, sampling from the sum of the two yields a positive value if $g(\vec{x}) > 0$. Additionally, $g(\vec{x})$ is positive so long as the probability of a positive sample is greater than $50\%$, but there is no need for this to be true for the method to work. Indeed, if there is a $0\%$ probability of a positive sample then $g(\vec{x}) = -1$, yielding the expected $-\textnormal{PDF}_-(\vec{x})$.

The use of this broadly applicable formalism to MC collider physics events generated is straightforward. The first step is to select a minimal set of variables to define the feature space, with the simultaneous goal of ensuring that the resulting reweighting remains maximally transferable to higher-level observables used in downstream analyses. In the context of collider physics, a natural choice for this feature space includes all relevant incoming and outgoing kinematic variables, upon which final-state kinematics are expected to depend. If, during validation, a particular observable fails to be accurately reproduced, this signals that the chosen feature space lacks sufficient information to model that observable. In such cases, the feature must depend on additional variables not included in the original selection. The input variables could be expanded then. On the other hand, one could instead perform reweighting directly on high-level analysis-specific variables. Such an approach would be valid, and perhaps more optimal, for that particular analysis, but it would likely not generalize to others.

\subsection{\label{sec:application_strategy}Application strategy}

To understand the practical implementation of the reweighting, suppose a set of events is generated across some feature space $X$ corresponding to all possible values for $\vec{x}$. Of these events, $N_+$ are positively weighted and $N_-$ are negatively weighted. Let $Y$ be a region determined by some arbitrary but fixed selections. These selections need not be directly applied to $X$ and may result in the union of many unconnected regions in $X$. 

In region $Y$, there is a set of $M_+$ events with positive weights and $M_-$ events with negative weights. Each event has an associated weight, $w_n$, and feature vector, $\vec{x_n}$. 
%The weights can be distinct for all events.

To calculate the cross section $\sigma$ of the bin, the weights are summed with the signs assigned by the generator, giving
\begin{equation}
\sigma(Y) = \sum_{n\in M_+}w_n -\sum_{n\in M_-}w_n.
\end{equation}
When calculating $\sigprime$, each weight is multiplied by $g(\vec{x_n})$ to get
\begin{equation}
\sigprime(Y) = \sum_{n\in M_+\cup M_-}w_ng(\vec{x_n}).
\end{equation}

The difference used in $\sigma$ is turned into the scaled sum in $\sigprime.$ The magnitude of the terms in the sum will be smaller than those in the difference, so the uncertainty is expected to be lower. This will be quantified in the following section. If $g(\vec{x})$ cannot be known exactly, the goal is to approximate it as closely as possible (via machine learning or another method), while introducing systematic uncertainties that are large enough to cover any deviations, but small enough to still lower the net uncertainty.

\section{\label{sec:reweightnoUncert}Reweighting When $g(\vec{x})$ Is Known Exactly}

The simplest use of the reweighting is when $g(\vec{x})$ is known exactly. While this is expected to be a rare occurrence in practice, it provides a clean way to verify the reweighting method. Here, the increase in statistical precision for this case is derived and quantified. It is then demonstrated in practice with an example from quantum mechanics.

To begin, the uncertainty prior to and following the application of reweighting will be derived. Let bin $Y$ have an arbitrary but fixed selection as discussed in the previous section. Consider a counting experiment performed on it. To determine the expectation value and variance on the bin, sub-bins are made such that each MC event falls into its own bin. These are all modeled as independent Poisson variables with mean and standard deviation equal to the weight of the event. 

Designating a Poisson variable with mean 1 as $\poisson$, the expectation value for the cross section of the bin is
\begin{equation}
\mathbb{E}[\sigma(Y)] = \sum_{n\in M_+}w_{n}\mathbb{E}[\poisson] -\sum_{n\in M_-}w_{n}\mathbb{E}[\poisson] = \sum_{n\in M_+}w_{n} -\sum_{n\in M_-}w_{n}.
\end{equation}
Similarly, assuming that all the distributions are independent, the variance is
\begin{align}
\Var[\sigma(Y)] &= \sum_{n\in M_+}w_{n}^2\Var[\poisson] + \sum_{n\in M_-}w^2_{n}\Var[\poisson] \notag \\
&= \sum_{n\in M_+}w^2_{n} + \sum_{n\in M_-}w^2_{n} = \sum_{n\in M_+ \cup M_-}w_{n}^2.
\end{align}

Thus, the variance is the squared sum of the weights for the variance and the expectation value is the sum of the weights.

Applying this calculation to $\sigprime$ when $g$ is known precisely is straightforward as $g(\vec{x_n})$ is a constant for each $x_n$; it becomes part of the weight. This gives
\begin{equation}
\mathbb{E}[\sigprime(Y)] = \sum_{n\in M_+\cup M_-}w_{n}g(\vec{x_n})\mathbb{E}[\poisson]  = \sum_{n\in M_+\cup M_-}w_{n}g(\vec{x_n}) 
\end{equation}
and
\begin{align}
\Var[\sigprime(Y)] &= \sum_{n\in M_+ \cup M_-}w_{n}^2g(\vec{x_n})^2\Var[\poisson] \notag \\
&= \sum_{n\in M_+ \cup M_-}w_{n}^2g(\vec{x_n})^2 \leq \sum_{n\in M_+ \cup M_-}w_{n}^2 = \Var[\sigma(Y)].
\label{eqn:ideal}
\end{align}

As $|g(\vec{x})|\leq1$, the net variance is always reduced unless $g(\vec{x})$ is strictly $\pm 1$, in which case the variance does not change.

\subsection{\label{sec:double_slit}Double slit interference MC}
\begin{figure}[h!]
    \centering
    \begin{tikzpicture}[thick,scale=1.2]

        % Vertical axis (dashed, with arrows on both ends)
        \draw[<->] (0,-2.7) -- (0,2.7) node[above right] {$x$};

        % Slits: mask out parts of the axis
        \draw[white,line width=2pt] (0,-1.0) -- (0,-0.5);
        \draw[white,line width=2pt] (0,0.5) -- (0,1.0);

        % Indicate slit size delta on the left
        \draw[<->] (-0.4,-1.0) -- (-0.4,-0.5);
        \node[left] at (-0.4,-0.75) {$\delta$};

        \draw[<->] (-0.4,0.5) -- (-0.4,1.0);
        \node[left] at (-0.4,0.75) {$\delta$};

        % Label slit positions alpha and -alpha on the right
        \node[right] at (0.4,0.75) {$\alpha$};
        \node[right] at (0.4,-0.75) {$-\alpha$};

        % Marker for y=0
        \fill (0,0) circle (1pt) node[right] {};
        \node[right] at (0.4,0) {$0$};

    \end{tikzpicture}
    \caption{Double-slit barrier with slits of width $\delta=0.25$ at positions $\pm\alpha=1$. }
    \label{fig:double-slit}
\end{figure}
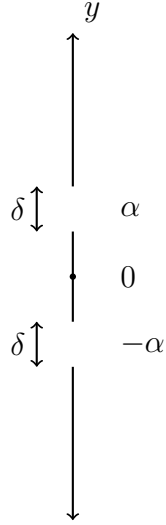

This prediction that reweighting lowers the statistical uncertainty of existing samples will now be demonstrated in practice. The double slit experiment is used for this purpose. For the sake of simplicity, the wave function of the outgoing photon in momentum space is used instead of the observed interference pattern, as are the natural units with $\hbar = c = 1$. The setup consists of two slits at $\alpha$ and $-\alpha$, each of width $\delta$, as shown in Figure~\ref{fig:double-slit}. Values of $\alpha=1$ and $\delta=0.25$ are used. An equal amount of light from far away sources is assumed to travel through each. 

The wave function is then
\begin{equation}
|\Psi\rangle = \frac{1}{\sqrt{4\delta}}\left(\int_{\alpha-\delta}^{\alpha+\delta}|x\rangle dx + \int_{-\alpha-\delta}^{-\alpha+\delta}|x\rangle dx\right)
\end{equation}
in position space and
\begin{equation}
\begin{aligned}
\langle p|\Psi\rangle &= \frac{1}{\sqrt{4\delta}}\left(\int_{\alpha-\delta}^{\alpha+\delta}\langle p|x\rangle dx + \int_{-\alpha-\delta}^{-\alpha+\delta}\langle p|x\rangle dx\right) \\
&= \frac{1}{\sqrt{8\pi\delta}}\left(\int_{\alpha-\delta}^{\alpha+\delta}e^{-ipx} dx + \int_{-\alpha-\delta}^{-\alpha+\delta}e^{-ipx} dx\right)
\end{aligned}
\end{equation}
in momentum space.
Performing a change of variables in the second integral from $x$ to $-x$ and a flip of the integration bounds introduces one minus sign from each operation, changes the sign of the exponential, and changes the integration bounds to those of the first integral. This yields
\begin{equation}
\langle p|\Psi\rangle = \frac{1}{\sqrt{8\pi\delta}}\int_{\alpha-\delta}^{\alpha+\delta}\left(e^{-ipx} + e^{ipx}\right) dx  = \frac{1}{\sqrt{2\pi\delta}}\int_{\alpha-\delta}^{\alpha+\delta} \cos(px)dx. 
\end{equation}

Integrating with respect to $x$ gives
\begin{equation}
\langle p|\Psi\rangle = \frac{1}{p\sqrt{2\pi\delta}}\left(\sin(p(\alpha+\delta)) - \sin(p(\alpha -\delta))\right). 
\end{equation}
Squaring to obtain the probability yields 
\begin{equation}
P(p) = |\langle p|\Psi\rangle|^2 = \frac{1}{2\pi p^2\delta}\left[\sin(p(\alpha+\delta)) - \sin(p(\alpha-\delta)) \right]^2.
\end{equation}

\begin{figure}[h]
    \centering
    \includegraphics[width=0.4\textwidth]{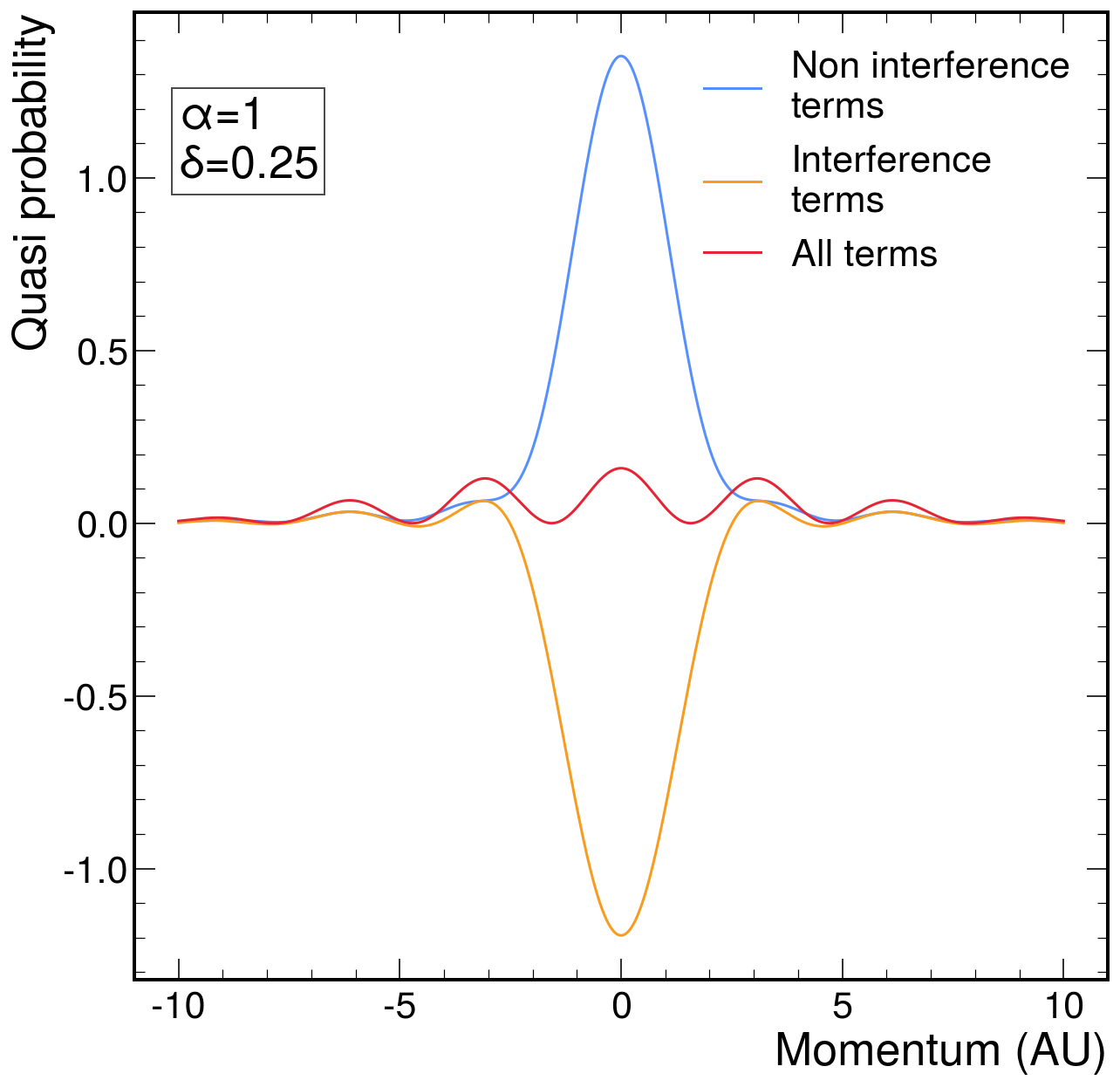}
    \includegraphics[width=0.4\textwidth]{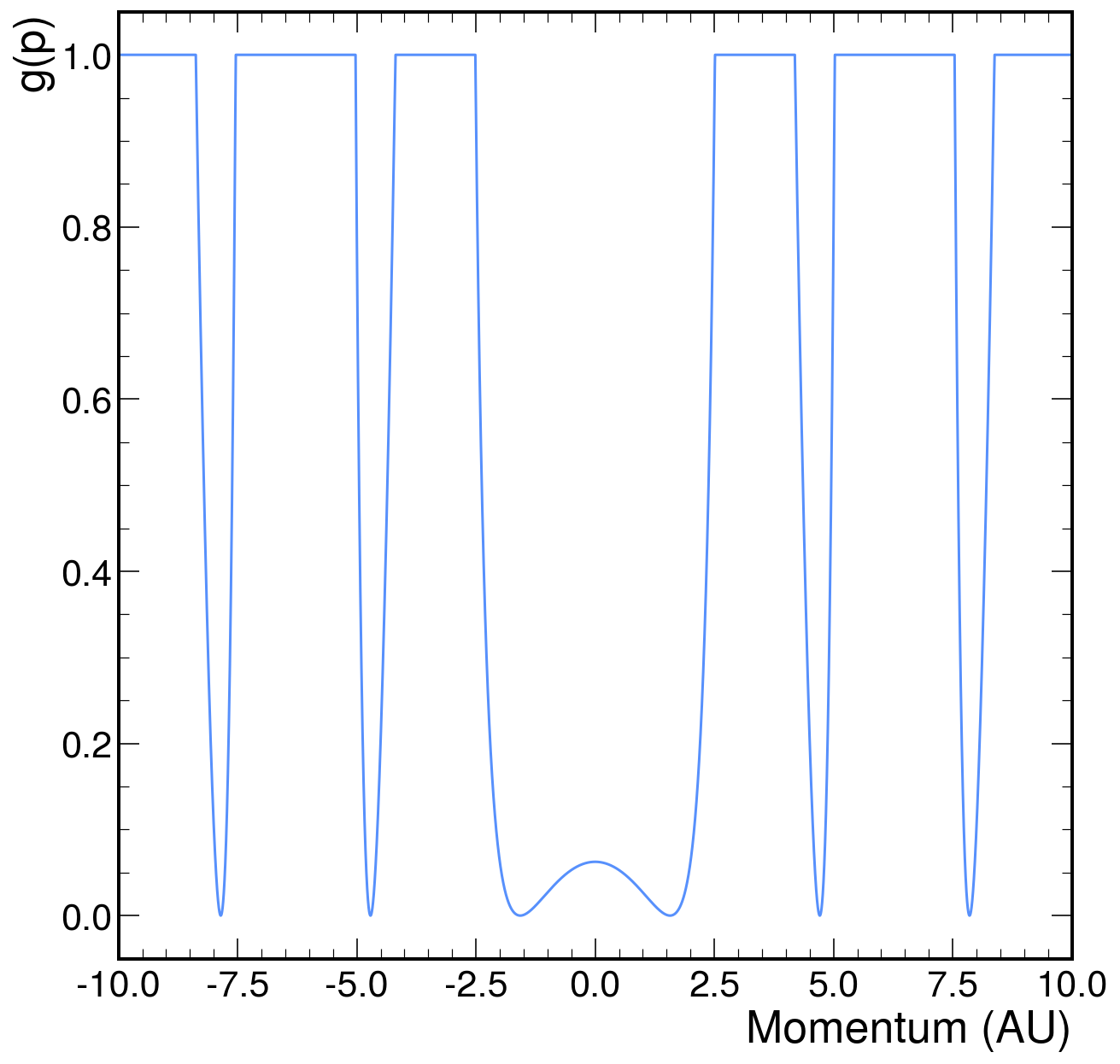}
    \caption{For the left plot, base terms are in blue and the interference terms are in orange. There is a lot of cancellation between the two near a momentum of 0, leading to the much smaller combined term. The right plot shows the reweighting function at each value of $p.$}
    \label{fig:qm_interference}
\end{figure}

This equation is now divided into two terms for the purposes of MC sampling. First, the always-positive noninterference term is
\begin{equation}
P_{\textnormal{base}}(p)= \frac{1}{2\pi p^2\delta}\sin^2(p(\alpha+\delta)) + \sin^2(p(\alpha-\delta)).
\end{equation}
Second, the positive or negative interference term is
\begin{equation}
P_{\textnormal{interference}}(p) = -\frac{1}{\pi p^2\delta}\sin(p(\alpha+\delta))\sin(p(\alpha-\delta)).
\end{equation}
These functions are shown in the left plot of Figure~\ref{fig:qm_interference}.

Using rejection sampling, MC samples from these distributions are produced in several steps. To start, bounds on the momentum and PDF values are set and pairs of numbers are uniformly generated within those ranges. Each pair is accepted if the PDF coordinate is less than the PDF value for the momentum coordinate. Each sample is weighted by the area of the pair generation rectangle over the total number of points generated. Due to this normalization, all the samples can be combined together to give an MC approximation of the combined PDF. 

Three sets of samples are generated: the noninterference term, the positive part of the interference term, and the negative part of the interference term. Sampling the positive part of the interference term is functionally equivalent to the noninterference term, as the negative regions are never accepted. Sampling from the negative part is accomplished by multiplying the PDF by $-1$ and giving the sampled weights a negative sign. The precise parameters used are given in Table~\ref{tab:double_slit_params}. The number of samples is chosen to give all events the same absolute weight.

\begin{table}[h!]
\begin{center}
\resizebox*{0.8\textwidth}{!}{
\begin{tabular}{|l|l|l|l|}
\hline
Sample & Momentum range & PDF range & Number of samples \\
\hline
Noninterference & [$-$10,10] & [0, 1.5] & 100,000\\
Positive interference & [$-$10,10] & [0, 0.075] & 5,000\\
Negative interference & [$-$10,10] & [0, 1.5] & 100,000\\
\hline

\end{tabular}}
\caption{Double slit sampling parameters.}
\label{tab:double_slit_params}
\end{center}
\end{table}

The ratio of the interference term to the summed term is taken to determine the analytic reweighting, which is given by
\begin{equation}
P_{+}(p) = \textnormal{Min}\left(\frac{\sin^2(p(\alpha+\delta)) + \sin^2(p(\alpha-\delta))}{[\sin(p(\alpha+\delta)) + \sin(p(\alpha-\delta))]^2}, 1\right).
\end{equation}
 Plugging this into the formula for the reweighting function gives
\begin{equation}
g(p) = \textnormal{Min}\left(2\frac{\sin^2(p(\alpha+\delta)) + \sin^2(p(\alpha-\delta))}{[\sin(p(\alpha+\delta)) + \sin(p(\alpha-\delta))]^2}- 1, 1\right).
\end{equation}
This is shown in the right plot of Figure~\ref{fig:qm_interference}.

\begin{figure}[h]
    \centering
    \includegraphics[width=0.4\textwidth, valign=m]{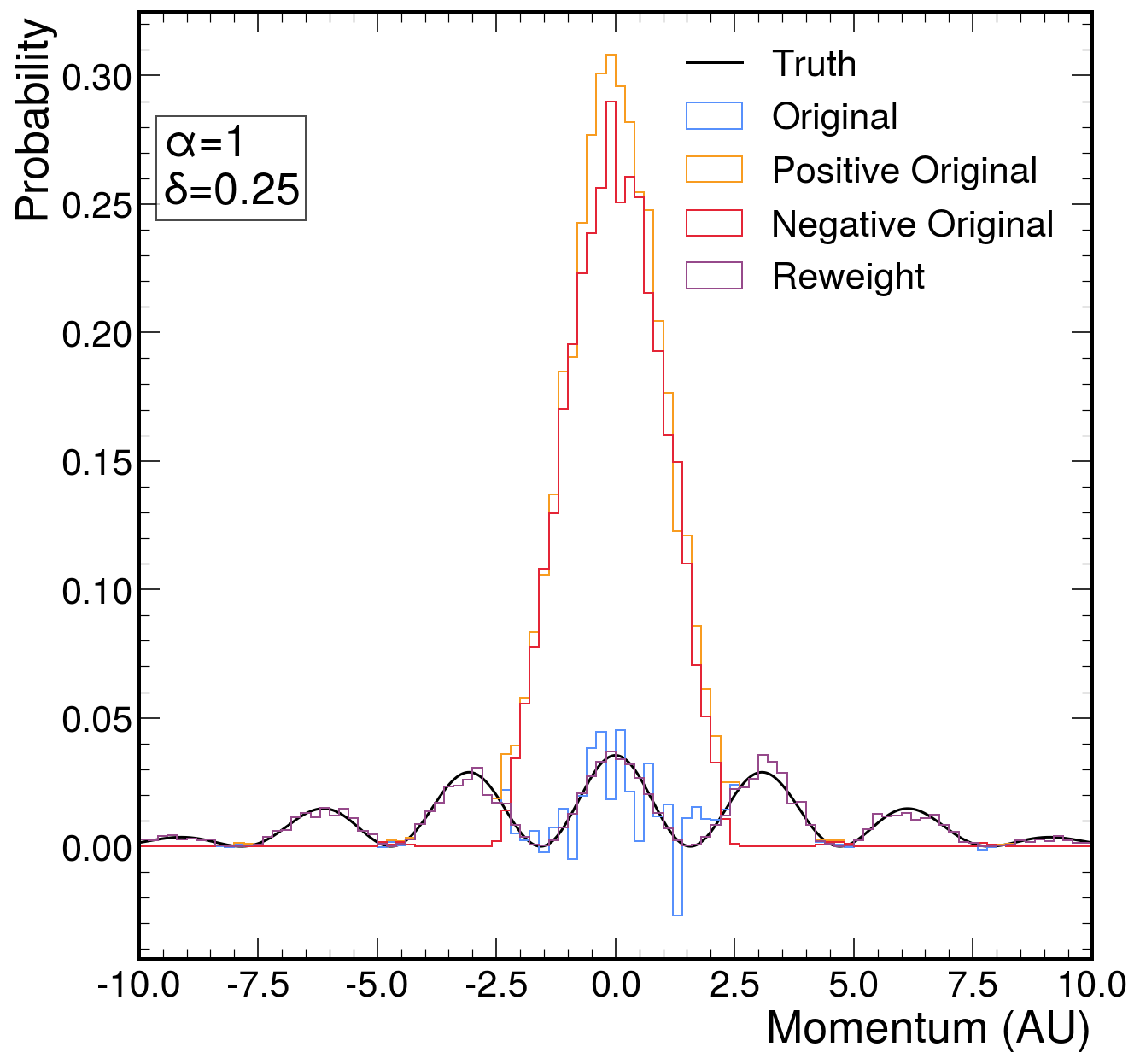}
    \includegraphics[width=0.4\textwidth, valign=m]{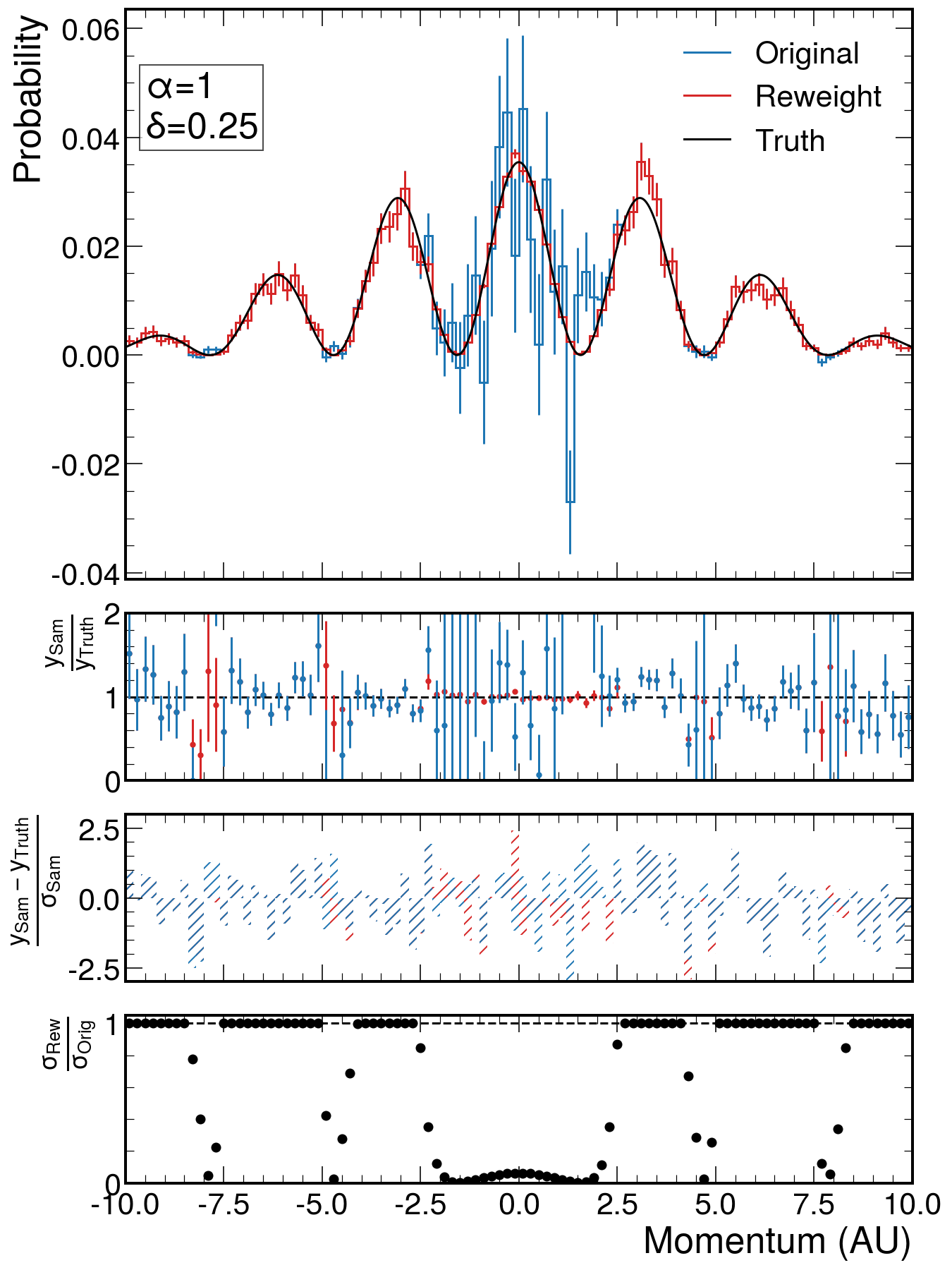}
    \caption{The left plot shows the true distribution, the originally sampled distribution, the reweighted distribution, and the positively and negatively weighted events. The right plot has a direct comparison of the true full distribution, the originally weighted distribution, and the reweighted distribution. The top panel compares their histograms, the second panel compares the ratios of the sampled histograms to the true values, the third panel shows the pull on the bin-by-bin MC statistical uncertainties give the uncertainty in each bin and the deviation from the truth, and the bottom panel shows the ratio of the uncertainties of the reweighted bins to the originals. Here $y$ refers to the bin value variable.}
    \label{fig:qm_histogram}
\end{figure}

Figure~\ref{fig:qm_histogram} shows the nominal and reweighted MC samples in histograms of momentum. The left plot shows the true function, the original and reweighted histograms, and the positive and negative components of the original histogram. The reweighted distribution agrees with the true distribution and has dramatically reduced stochasticity compared to the original.

The original, reweighted, and true histograms are shown in isolation in the first panel of the right plot, further demonstrating the closure and decreased stochasticity of the reweighted distribution. The second panel shows the ratios of the two histograms to the truth. In the regions with cancellation, the reweighted histogram ratio is much closer to 1 than the original histogram, while in the regions with no cancellation the two ratios are identical. The error bars on the points are of appropriate size, generally covering the distance between the point and one for both histograms. This is quantified in the third panel, where the pulls of the bin-by-bin uncertainties are plotted. The pulls are defined as difference between the original and reweighted histograms divided by the uncertainty of the difference. The pulls on the original and reweighted distributions are of similar size everywhere, including the regions where the uncertainties on the reweighted distribution are much smaller. Those regions are visible in the bottom panel where the ratio of the bin-by-bin uncertainties is shown. This shows that reweighting lowers the MC statistical uncertainties and that there is no need to inflate the uncertainties or resample the distributions to obtain appropriate uncertainties. 

\begin{figure}[h]
    \centering
    \includegraphics[width=0.3\textwidth, valign=m]{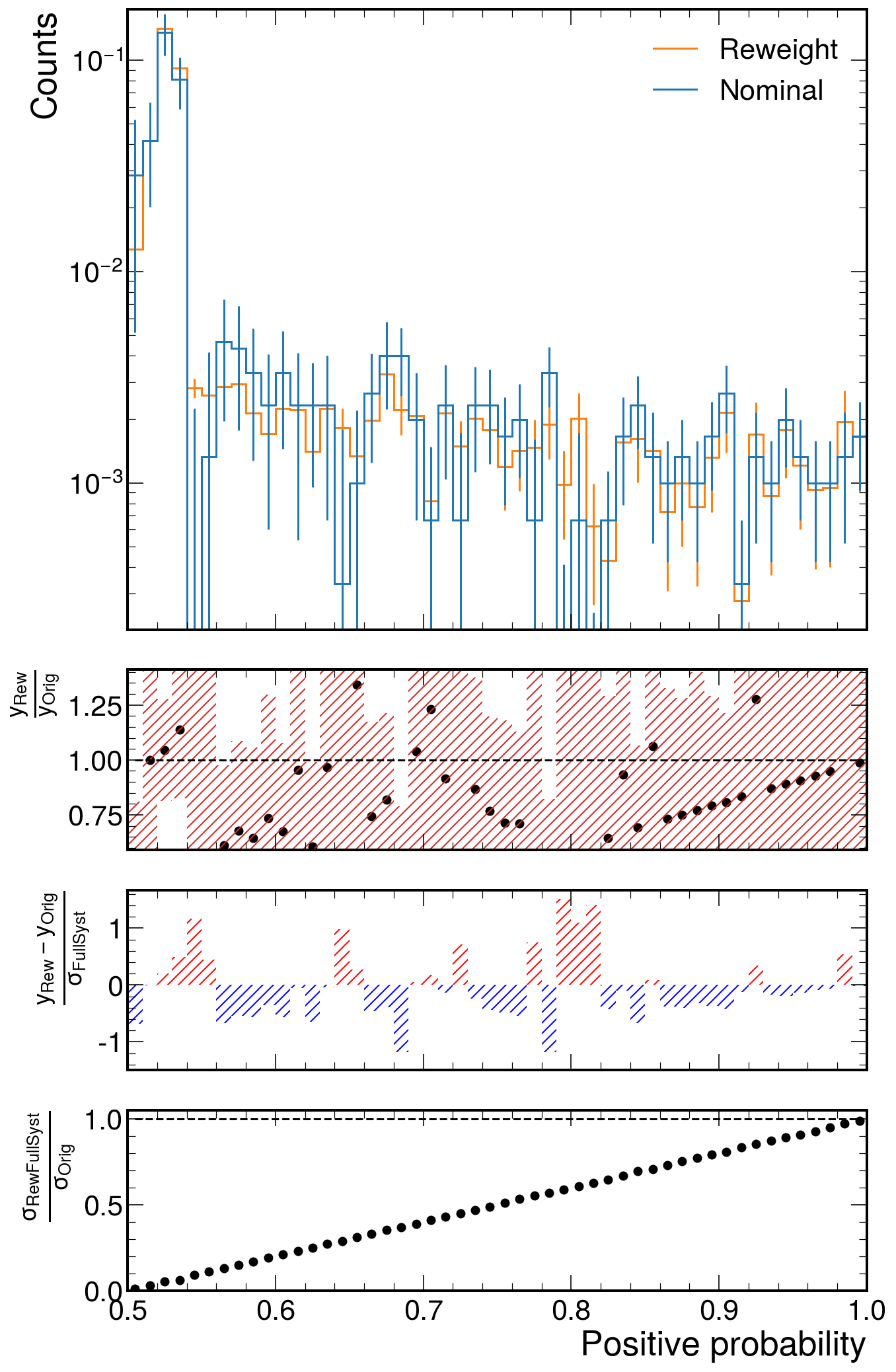}
    \includegraphics[width=0.3\textwidth, valign=m]{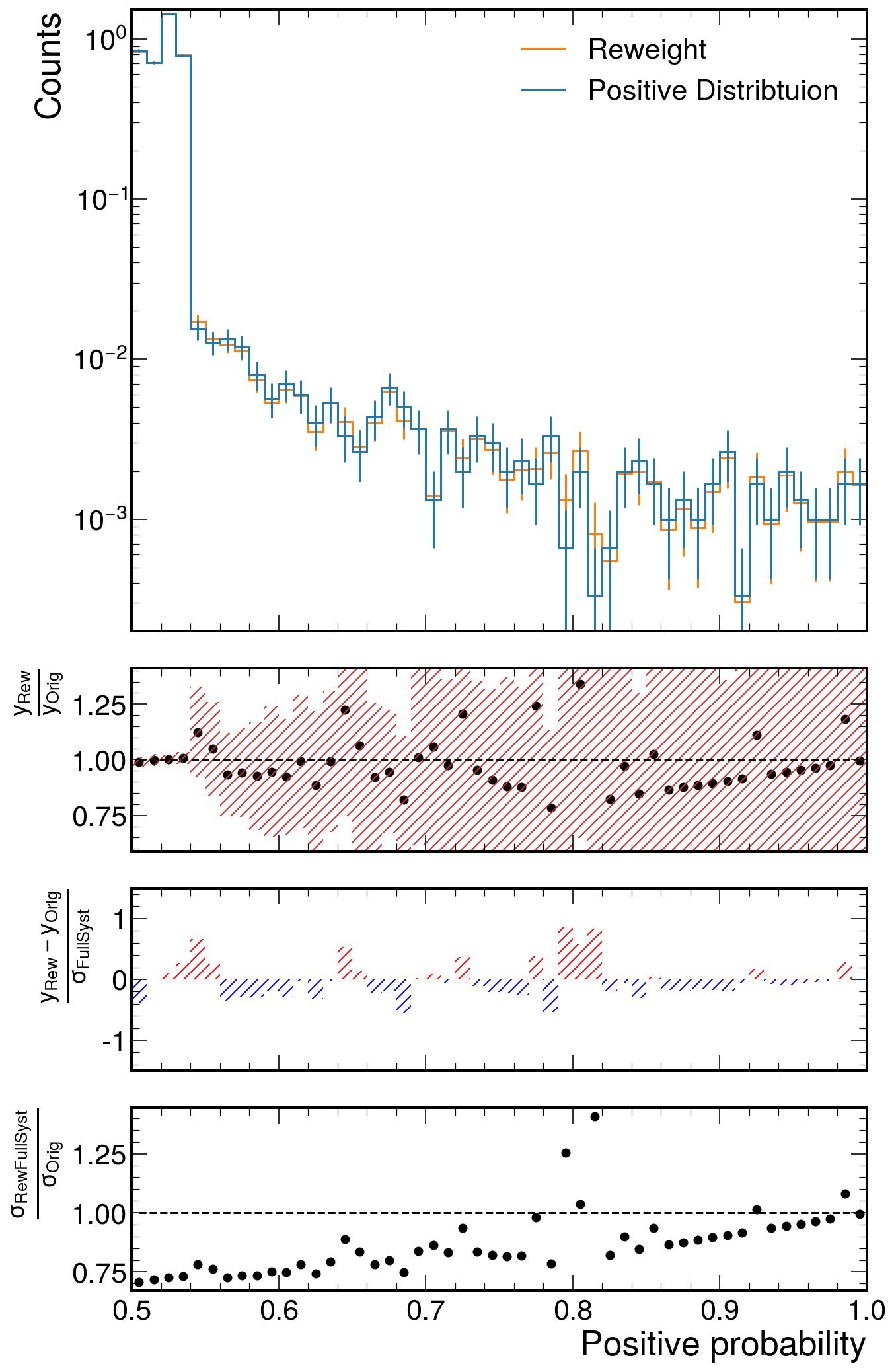}
    \includegraphics[width=0.3\textwidth, valign=m]{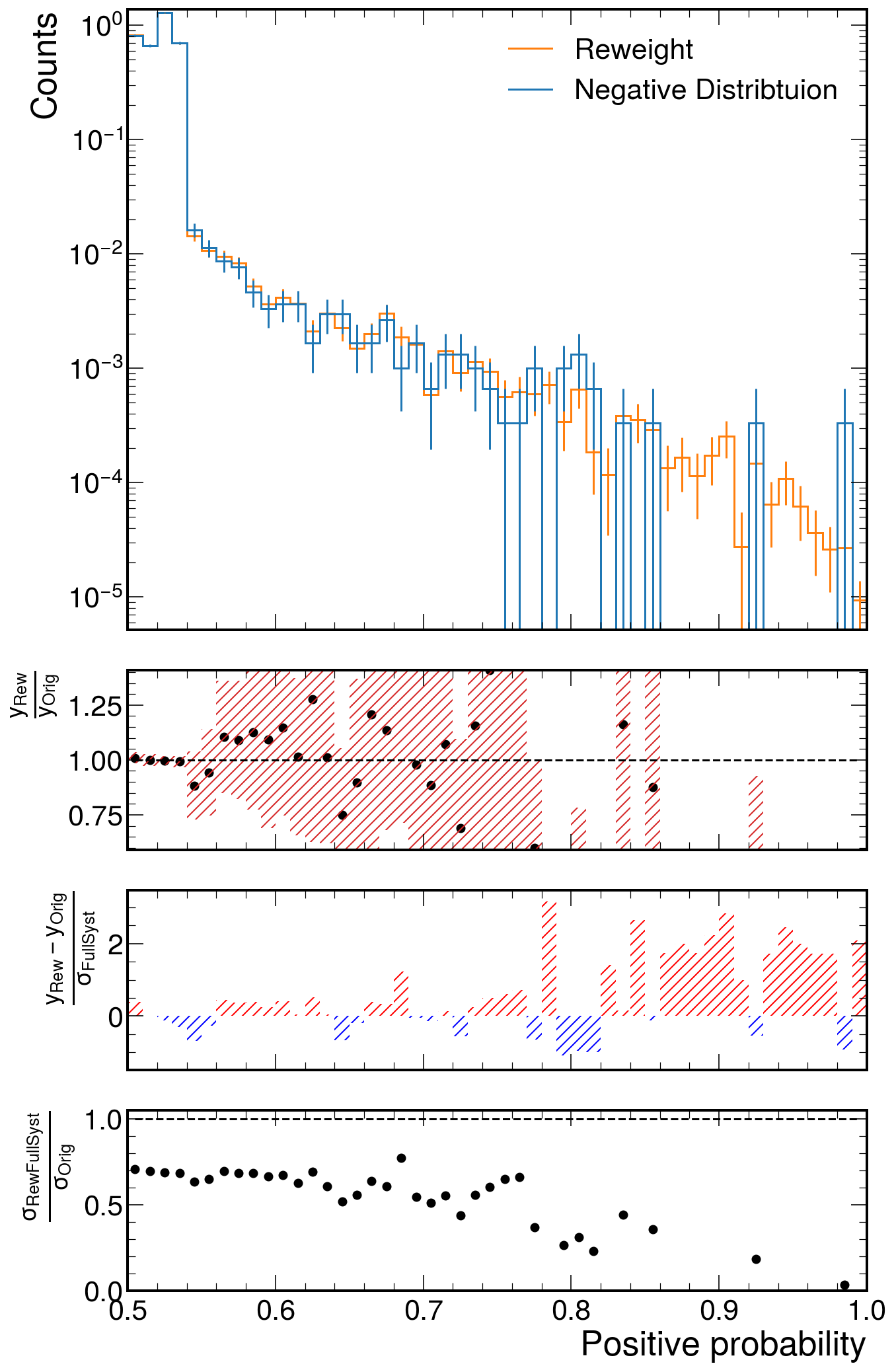}
    \caption{Closure on reweighting of the probability of being positive to the nominal distribution (left), the positive distribution (middle), and the negative distribution (right).}
    \label{fig:qm_closure}
\end{figure}

An additional test which can be performed is to reweight to the distributions of the positive or negative weights. Instead of multiplying by $g(x)$, one can multiply by $P_+(\vec{x})$ or $P_-(\vec{x})$. The result of applying the reweighting to the distribution of the probability of being positive is shown in Figure~\ref{fig:qm_closure}. The leftmost plot shows the proper distribution for the probability of being positive. The center plot shows the distribution the positively weight events follow, and the right plot shows the distribution the negatively weighted events follow. In all three cases, the reweighted sample histograms fall within the statistical uncertainty bounds of the nominally weighted histograms.

\section{\label{sec:uncertainty_quantification}Uncertainty Quantification of Reweighting}

The reweighting works very well when $g(\vec{x})$ is known precisely, but in most cases $g(\vec{x})$ is unknown and not analytically derivable. In these situations $g(\vec{x})$ must be approximated, and an appropriate uncertainty quantification must be included. In this paper a deep neural network (DNN) is used for this purpose. A direct functional fit could be performed for $g(\vec{x})$ given sufficient prior information, but no specific functional form is known \textit{a priori} here. The following text explains how a DNN can model $g(\vec{x})$, derives the general properties of the uncertainty, and presents two methods for quantifying and applying the uncertainty.

\subsection{\label{sec:model_choice}Model choice}

Modeling $g(\vec{x})$ through its relationship with $P_+(\vec{x})$ is straightforward as $g(\vec{x})$ is just $2P_+(\vec{x}) - 1$. To determine a suitable model, consider a binary classification function that minimizes the binary cross-entropy loss function on an infinite set of data. This model can be shown to model $P_+(\vec{x})$ exactly \cite{Gneiting2007}. Further, according to the Universal Approximation Theorem \cite{hornik1989multilayer}, an infinitely wide DNN can fit any function arbitrarily well. Thus, a DNN trained to classify the positively and negatively weighted events, \dnnre, is a good choice to model $P_+(\vec{x})$, and in turn $g(\vec{x})$.

To perform the uncertainty analysis, 20 different DNNs are trained using subsampling without replacement. For single predictions, the mean value of the 20 networks is used. This is equivalent to making 20 different histograms and taking the mean value for the bin, as both methods sum the weights of all 20 networks for each event in the bin and divide by 20. The only difference is when the division occurs: before building histograms when using the mean weight value or after building histograms when using the mean bin value.

The choice of a DNN here is not especially important; the following analysis holds for any equally flexible model  (e.g., a boosted decision tree, graph neural network, or multi-dimensional parametrization). The ability of the model to fit the desired functions well and have multiple configurations minimizing, or nearly minimize, the loss is the only relevant characteristic. The set of models will then have a distribution of predictions when interpolating or extrapolating, which is necessary for the uncertainty quantification.

In the case of an unbalanced dataset, it may be desirable to balance the number of positive and negative events in the training set. This can be done, as long as the \dnnre is rescaled afterward. For example, suppose that only a fraction $a$ of the original positively sampled events are used. Then the model will actually learn
\begin{equation}
P'_+(\vec{x}) = \frac{a\textnormal{PDF}_+(\vec{x}) }{a\textnormal{PDF}_+(\vec{x})+ \textnormal{PDF}_-(\vec{x})}.
\label{eqn:pplusprime}
\end{equation}
while the desired output is
\begin{equation}
P_+(\vec{x}) = \frac{\textnormal{PDF}_+(\vec{x}) }{\textnormal{PDF}_+(\vec{x})+ \textnormal{PDF}_-(\vec{x})}.
\label{eqn:pplus}
\end{equation}
An expression for the negative PDF can be extracted from Eq.~(\ref{eqn:pplus}).
\begin{equation}
\textnormal{PDF}_-(\vec{x}) = \textnormal{PDF}_+(\vec{x})\left(\frac{1}{P_+(\vec{x})} - 1\right)
\end{equation}
Substituting this into Eq.~(\ref{eqn:pplusprime}) gives
\begin{equation}
P'_+(\vec{x}) = \frac{a\textnormal{PDF}_+(\vec{x}) }{a\textnormal{PDF}_+(\vec{x})+ \textnormal{PDF}_+(\vec{x})\left(\frac{1}{P_+(\vec{x})} - 1\right)}.
\end{equation}
From here, some algebra yields
\begin{equation}
P_+(\vec{x}) = \frac{1}{1+ a\frac{1-P'_+(\vec{x})}{P'_+(\vec{x})} }.
\end{equation}
Thus, as long as the output of the model is transformed according to this function, the model can be trained using a rebalanced dataset. 

\subsection{\label{sec:uncertainty_properties}Properties of uncertainty}

In many analyses, the goal is to measure $\frac{d\sigma}{dX}$, where $X$ is some kinematic or identification variable. This goal is achieved through histograms of observables related to $\frac{d\sigma}{dX}$ and constrainable systematic uncertainties. These distributions are often used in a fit to extract $\frac{d\sigma}{dX}$ from observed data. Thus, a means of propagating an uncertainty on the modeling of $g(\vec{x})$ to these histograms in a binwise manner is needed. 

This uncertainty has a statistical (aleatoric) component reflecting the statistical fluctuations in the training data used for each of the models.
It also has a systematic (epistemic) modeling uncertainty due to the finite size of the \dnnre and the imperfections in the gradient descent algorithms used to train it.

The uncertainty in the model is treated like a systematic uncertainty as commonly used in HEP. The most accurate way to determine the uncertainty is to apply each of the models separately and obtain a collection of different histograms at the end of the analysis. This ensures that all of the available information is kept until the end of the analysis, allowing the optimal uncertainties to be assigned. It is typical, however, in HEP analyses to use simple up and down uncertainties that are applied per event. This approach, while simple, removes much of the available information. In particular, it distills the full uncertainty distribution to a center value and two variations and removes most of the information on the correlation between events beyond the simple up and down correlation. This is sufficient for a mostly gaussian uncertainly with a high degree of correlation, but is not suitable for distributions with any additional complexity.  A method for applying the uncertainty using both of these methods is presented in the following sections, but it will be clear that the histogram based approach provides far more accurate uncertainties. 

The event-based approach to the uncertainty is the simplest and is similar to many other HEP uncertainties. Up and down variations are calculated for $g(\vec{x})$, applied to each MC event, and propagated to the final histograms. The histogram-based uncertainty method uses alternative final histograms to obtain systematic shape uncertainties with principal component analysis (PCA). This method requires additional work after making the histograms, but easily accounts for the covariance of individual predictions and produces meaningful shape uncertainties. The PCA method is very robust to the expected covariances between the predicted distributions of $g(\vec{x})$ at different $\vec{x}$ \cite{Jolliffe2002PCA}, while they are simply ignored by the event-based approach.

\begin{comment}

A complicating factor in both approaches is the covariance between the predicted distributions of $g(\vec{x})$ at different $\vec{x}.$ These correlations arise because each distribution is determined by the same alternate models of $g(\vec{x})$. If a given fit predicts a higher value at $\vec{x_0}$, it may also predict a higher (or lower) value at $\vec{x_1}$, depending on how the alternate models co-vary. Consequentially, the distributions at different $\vec{x}$ are correlated, with the strength and sign of the correlation determined by how uniformly the models vary across $\vec{x}$.

\begin{figure}[h]
    \centering
    \includegraphics[width=0.6\textwidth, valign=m]{covariance.png}
    \caption{An example of two possible alternative shapes.}
    \label{fig:covariance}
\end{figure}

Suppose there are two alternative models for $g(\vec{x})$ from separate trainings on a 1-D feature space, such as the orange and blue curves in Figure~\ref{fig:covariance}. When these shapes are used in a signal-extraction fit, the final result moves towards one or the other of the histograms. If the result moves towards the orange to bring the mean up near 1, the mean also increases for positive values of the feature space, but decreased for negative values. Thus, there is a correlation at these points, meaning there is a nonzero covariance between the distributions. This also happens with the reweighting function, except there are 20 alternative shapes from the 20 fits, not 2, making the correlations more complex.

\end{comment}

\subsection{\label{sec:event_uncertainties}Event-by-event uncertainties}

To obtain the event-by-event uncertainties, the same approach as in Sec.~\ref{sec:reweightnoUncert} is used, except that each event weight is also multiplied by $g(\vec{x_n})$, which is assumed to have its own distribution for each $\vec{x_n}.$ Further, nonzero covariance between these distributions is assumed. Let $N_{\mathrm{tot}}$ be the total number of MC events in the bin. The expectation value for the reweighted cross section is then
\begin{equation}
\mathbb{E}[\sigprime(Y)] = \sum_{n\in M_{+}\cup M_{-}}w_{n}\mathbb{E}[g(\vec{x_n})]\mathbb{E}[\mathds{1}].
\end{equation}

Obtaining the variance on the cross section requires the consideration of the correlation between $g(\vec{x_n})$ for different $\vec{x_n}$. This makes the full formula
\begin{equation}
\Var(\sigprime(Y)) = \sum_{n}^{N_{\mathrm{tot}}}\sum_{m}^{N_{\mathrm{tot}}}\Cov(w_{n}g(\vec{x_n})\mathds{1}, w_{m}g(\vec{x_m})\mathds{1}).
\end{equation}
The constant weights are pulled out, and the sum is broken into two categories, one with $n=m$ and one with $n\neq m.$ This gives
\begin{equation}
\begin{aligned}
\Var(\sigprime(Y)) &= \sum_{n}^{N_{\mathrm{tot}}}w_{n}^2\Var(g(\vec{x_n})\mathds{1})  \\
&+ \sum_{n\neq m}^{N_{\mathrm{tot}}}w_{n}w_{m}\Cov(g(\vec{x_n})\mathds{1},g(\vec{x_m})\mathds{1}).
\end{aligned}
\end{equation}
In the diagonal term $g(\vec{x})$ and the Poisson distributions are independent. This gives

\begin{equation}
\begin{aligned}
&\sum_{n}^{N_{\mathrm{tot}}}w_{n}^2(\mathbb{E}[g(\vec{x_n})]^2\Var(\mathds{1}) + \Var(g(\vec{x_n}))\mathbb{E}[\mathds{1}]^2 +  \Var(\mathds{1})\Var(g(\vec{x_n})))  =\\
&\sum_{n}^{N_{\mathrm{tot}}}w_{n}^2(g(\vec{x_n})^2 + 2(\delta g(\vec{x_n}))^2)
\end{aligned}
\end{equation}
where $\delta^2(g(\vec{x_n}))$ is the variance of $g.$

The off diagonal term is more complex as the covariances on $g$ are difficult to know \textit{a priori}, given their dependence on the model used and how it is trained. The Poisson distributions are assumed to be independent of each other and the other functions, so they are dropped. Replacing the covariance with the more interpretable correlation coefficient  $\rho$ times the standard deviations of the two variables, the off diagonal part becomes
 \begin{equation}
 \sum_{n\neq m}^{N_{\mathrm{tot}}}w_{n}w_{m}\delta g(\vec{x_n})\delta g(\vec{x_m})\rho(g(\vec{x_n}), g(\vec{x_m})).
 \end{equation}
This makes the full formula for the variance
\begin{equation}
\begin{aligned}
\Var(\sigprime(Y)) &= \sum_{n}^{N_{\mathrm{tot}}}w_{n}^2(g(\vec{x_n})^2 + 2(\delta g(\vec{x_n}))^2)  \\
&+ \sum_{n\neq m}^{N_{\mathrm{tot}}}w_{n}w_{m}\delta g(\vec{x_n})\delta g(\vec{x_m})\rho(g(\vec{x_n}), g(\vec{x_m})).
\end{aligned}
\end{equation}

Calculating this is impractical in most use cases, as it requires storing all entries into histogram bins and then carrying out an order $N^2$ algorithm on each of them. The simplified cases of an exactly known $g$, a $g$ with no covariance at different $\vec{x}$, and a $g$ with maximal covariance at different $\vec{x}$ are therefore examined to understand the behavior of the uncertainty in these limits.

First, suppose there is no uncertainty on $g$. This just leaves
\begin{equation}
\Var(\sigprime(Y)) = \sum_{n}^{N_{\mathrm{tot}}}w_{n}^2g(\vec{x_n})^2.
\end{equation}
This is the ideal case and matches what was derived in Sec.~\ref{sec:reweightnoUncert}. Assuming that $g$ is approximately the same for all $n$ in the sum, the ratio of the variances is just $g(\vec{x_n})^2$. Thus, the overall uncertainty ratio scales with $|g|$.

Next, the point-by-point correlations are taken to be 0, so the covariance of $g$ at all $\vec{x}$ is 0. The second term completely drops out, giving
\begin{equation}
\Var(\sigprime(Y)) = \sum_{n}^{N_{\mathrm{tot}}}w_{n}^2(g(\vec{x_n})^2 + 2(\delta g(\vec{x_n}))^2).
\end{equation}
This indicates that the variances decrease with reweighting so long as the uncertainty on $g$ is sufficiently small. It adds a penalty per sample for not knowing $g$ exactly. If the combination of the uncertainty and $g$ is less than 1, the net uncertainty will decrease.  This is not a realistic assumption for small bins, which likely have similar $\vec{x}$ and a high degree of correlation. Over a large bin, however, this is a more realistic assumption as the correlations will partially cancel. 

Finally, a significantly worse case is considered. If the weights are taken to all have the same value, $g$ is assumed to be fully correlated everywhere, and $g$ is assumed to have the same value everywhere, the variance is

\begin{equation}
\begin{aligned}
\Var(\sigprime(Y)) &= \sum_{n}^{N_{\mathrm{tot}}}w^2(g(\vec{x_n})^2 + 2(\delta g^2)) + \sum_{n\neq m}^{N_{\mathrm{tot}}}w^2(\delta g)^2 \\ 
&= (Nw\delta g)^2 + Nw^2(g^2 + \delta g^2)
\end{aligned}.
\end{equation}

The total number of events in the bin is simply $Nwg$ (assuming that $g$ is correct). This means the relative uncertainty is

\begin{equation}
 \frac{\sqrt{(Nw\delta g)^2 + Nw^2(g^2 + \delta g^2)}}{Nwg} = \sqrt{\left(\frac{\delta g}{g}\right)^2\left(1+\frac{1}{N}
 \right) + \frac{1}{N} }.  
\end{equation}
In contrast, the relative uncertainty with no reweighting is 
\begin{equation}
 \frac{\sqrt{Nw^2}}{Nwg} = \frac{1}{g\sqrt{N}}.  
\end{equation}

In the limit of large $N$ (regardless of the weight), most terms in both equations will be negligible, leaving the reweighted relative uncertainty as $\frac{\delta g}{g}$, while the original relative uncertainty goes to 0. Thus, there is a lower bound for the reweighted relative uncertainty that does not exist for the original uncertainty. When $N$ is small, the reweighted relative uncertainty is still smaller than the original. So long as 
\begin{equation}
 N < \frac{1-g^2-\delta g^2}{\delta g^2},  
 \label{eqn:uncertGain}
\end{equation}
the reweighted relative uncertainty will be lower. For example, with $g=0.7$ and $\delta g = 0.07$ this gives $N<103$.

Assuming full correlation for all the uncertainties in a bin is a poor assumption for large bins. However, for tightly selected events with a small feature space, full correlation may be reasonable. In finely binned, high signal-to-background regions, uniform event properties may justify assuming full correlation within a bin.

The equations above elucidate qualitative features that will be observed in forthcoming examples. In particular, Eq.~(\ref{eqn:uncertGain}) shows that given enough events, the original statistical precision can be smaller than the uncertainty in the model. Because histogram bins are statistically independent, some bins may show increased total uncertainty while others are reduced. However, in a typical physics search, the most signal-rich kinematic phasespace will suffer the most from limitations in MC sample sizes. Thus, those bins---typically the most significant bins---show reduced overall uncertainty (i.e., combined statistical and model systematic uncertainty). In some cases, the improvement can approach the ideal case of a reduction by the factor of $g(\vec{x_n})$, as shown in Eq.~(\ref{eqn:ideal}), when the statistical uncertainty is much larger than the model uncertainty.

\subsection{\label{sec:histogram_uncertainties}Final observable level uncertainties}

The alternative uncertainty propagation method that avoids the computational issues and flawed approximation from the previous section is to use the subsampled models to construct alternative histograms. These histograms are then used to calculate the systematic uncertainties directly for the bin values. This approach requires storing more histograms, but it makes the final covariance calculation feasible.

Each trained $g(\vec{x})$ estimator is used to fill the histograms used in a signal-extraction fit, and the values for the histogram bins are treated as random variables. As the number of histogram bins and alternative shapes is not large, the full covariance matrix between the different histogram bins is calculable, as well as its PCA vectors. The primary components are orthogonal by construction and conventionally ordered by decreasing variance. While the first few components contribute the bulk of the uncertainty, up to the number of estimators of systematic uncertainties per histogram are available. This approach provides more information than the event-by-event approach, even when the full covariance is calculated, as the event-by-event approach can only provide a single set of up and down variations.

\section{\label{sec:VH_example}HEP Example}

To show how to learn $g(\vec{x})$, how it performs on HEP MC samples, and the size of the PCA-based systematic uncertainties, the reweighting is performed on a set of vector boson plus jets, \vjets, samples. This demonstrates that $g(\vec{x})$ is learnable for real physics MC and can provide good closure. Next, the reweighted samples are used as the background in a search for a Higgs boson produced in association with a $Z$ boson, decaying to bottom quarks and neutrinos:  $ZH\to \nu\bar{\nu}b\bar{b}$. A signal histogram is made using a signal-versus-background DNN (\dnnsvb) trained to separate the $VH$ events from the \vjets events. As the most sensitive \dnnsvb bins are created from strict, nonlinear selections on a mixture of reconstruction-level variables, this procedure is ideal to validate this reweighting method. If the reweighting closes here, it is a strong indication of its ability to close for a wide variety of analysis needs. The measured reduction in statistical uncertainty in these sensitive bins will also provide an excellent demonstration of the method's power.

For simplicity, several other backgrounds are omitted, including $t\bar{t}$, single top, diboson (ZZ, WZ), and $W\ell\nu Hb\bar{b}$ processes. The included samples are still sufficient to demonstrate the efficacy of the reweighting.

\subsection{\label{sec:samples}Samples used}

The Sherpa samples used for the \vjets processes are in Table~\ref{tab:sherpa_samples}. The names listed are the base names of the samples, and each real sample has either \_BFilter, \_CFilterBVeto, or \_CVetoBVeto added. BFilter selects for any B hadrons, CFilter selects for D hadrons while Veto reject events passed by the corresponding filter. Together, these three sample types partition the full event space, ensuring that all relevant hadronic configurations are represented. The processes have vector boson production and decay plus 0, 1, or 2 jets at NLO or 3, 4, or 5 jets at LO using a merging scale of 20\,GeV. The Sherpa version used for all the samples is 2.2.11 except for the $Z\tau\tau$ samples, which used 2.2.14. 
The samples were processed to the PhysLite \cite{Schaarschmidt:2024PhysLite} format using Athena release 22 \cite{ATLAS_Athena_2019_Release22} as part of the ATLAS MC20a campaign and released on CERN OpenData \cite{CERN_OpenData_Portal} as part of the electroweak boson nominal samples. The cross sections used are from Ref.~\cite{atlas_evgen_metadata}.

\begin{table}[htbp]
\centering
\begin{tabular}{|l|l|l|}
\hline
Sample name & Process& xsec [pb]  \\
\hline
mc20\_13TeV\_MC\_Sh\_2211\_Zee\_maxHTpTV2 & $Zee$ + jets & 2220 \\
mc20\_13TeV\_MC\_Sh\_2211\_Zmumu\_maxHTpTV2 & $Z\mu\mu$ + jets & 2220 \\
mc20\_13TeV\_MC\_Sh\_2214\_Ztautau\_maxHTpTV2 & $Z\tau\tau$ + jets & 2240 \\
mc20\_13TeV\_MC\_Sh\_2211\_Znunu\_pTV2 & $Z\nu\nu$ + jets  & 447 \\
mc20\_13TeV\_MC\_Sh\_2211\_Wenu\_maxHTpTV2 & $We\nu$ + jets  & 21800\\
mc20\_13TeV\_MC\_Sh\_2211\_Wmunu\_maxHTpTV2 & $W\mu\nu$ + jets  & 21800 \\
mc20\_13TeV\_MC\_Sh\_2211\_Wtaunu\_L\_maxHTpTV2 & leptonic $W\tau\nu$ + jets  & 7700 \\
mc20\_13TeV\_MC\_Sh\_2211\_Wtaunu\_H\_maxHTpTV2 & hadronic $W\tau\nu$ + jets  & 14100\\
\hline

\end{tabular}
\caption{\vjets Sherpa samples.}
\label{tab:sherpa_samples}
\end{table}

The Powheg samples used for the Higgs boson signal samples are in Table~\ref{tab:powheg_samples}. They all have the Z boson decaying to two neutrinos and the Higgs boson decaying to two b quarks. The samples were generated using the Powheg $ggHZ$ and $HZ+1J$ processes. As with the Sherpa samples, the samples were processed to PhysLite format using Athena release 22 as part of the ATLAS MC20a campaign and released on CERN OpenData, but as part of the Higgs boson nominal samples. The cross sections used are also from Ref.~\cite{atlas_evgen_metadata}.

\begin{table}[htbp]
\centering
\begin{tabular}{|l|l|l|}
\hline
Sample name & Process & xsec [pb]  \\
\hline
\makecell{mc20\_13TeV\_MC\_PowhegPythia8EvtGen\_\\NNPDF3\_AZNLO\_ggZH125\_vvbb} & $ggZ\nu\nu Hbb$ & 0.01429 \\
\makecell{mc20\_13TeV\_MC\_PowhegPythia8EvtGen\_\\NNPDF3\_AZNLO\_ZH125J\_MINLO\_vvbb\_VpT} & $Z\nu\nu Hbb$ & 0.08866 \\
\hline
\end{tabular}
\caption{VH Powheg samples.}
\label{tab:powheg_samples}
\end{table}

The ATLAS recorded luminosity from Run2 of 140.1 fb$^{-1}$ \cite{ATLAS:2022hro} is used for the luminosity in the paper.

\subsection{\label{sec:model_training}Model training}

To approximate $g(\vec{x})$, 20 \dnnre are trained using subsampled sets of the training data. The training variables are chosen to be as close to the kinematics of the underlying hard scatter as possible. Reconstruction-level variables are usable, but using them will not guarantee closure on generator-level variables or other independent or semi-independent reconstruction-level variables. Including all of the kinematic information (for incoming and outgoing partons) that is sampled during MC generation should be sufficient, as this is the information present at the time the weight is assigned. Not all of this information is necessary, however, and can cause the \dnnre to learn false correlations \cite{geirhos2020shortcut}. Some variables can therefore be pruned. Adding relevant, derived features is also very helpful, even if the \dnnre can in theory learn those variables itself \cite{domingos2012few}. In particular, variables helping to establish the Feynman diagram that produced the event and variables describing the interaction with the parton shower can be very helpful.

\begin{table}[h!]
\begin{center}
\resizebox*{0.6\textwidth}{!}{
\begin{tabular}{|l|l|l|}
\hline
$\pt(l_1)$  & $\eta(l_1)$ & $\phi(l_1)$\\
$\pt(l_2)$  & $\eta(l_2)$ & $\phi(l_2)$ \\
$V$ flavor  &  &  \\
\hline
$\pt(p_1)$  & $\eta(p_1)$ & $\phi(p_1)$ \\
$\pt(p_2)$  & $\eta(p_2)$ & $\phi(p_2)$ \\
$\pt(p_3)$  & $\eta(p_3)$ & $\phi(p_3)$ \\
$\pt(p_4)$  & $\eta(p_4)$ & $\phi(p_4)$ \\
$\pt(p_5)$  & $\eta(p_5)$ & $\phi(p_5)$ \\
\hline
$k_T(p_1,p_2)$ & $k_T(p_1,p_3)$ & $k_T(p_1,p_4)$ \\ $k_T(p_1,p_5)$ & $k_T(p_2,p_3)$ & $k_T(p_2,p_4)$ \\ $k_T(p_2,p_5)$ & $k_T(p_3,p_4)$ & $k_T(p_3,p_5)$ \\$k_T(p_4,p_5)$ & &\\
\hline
$k_T^{\text{PS}}(1,0)$ & $k_T^{\text{PS}}(2,1)$ & $k_T^{\text{PS}}(3,2)$ \\
$k_T^{\text{PS}}(4,3)$ & $k_T^{\text{PS}}(5,4)$ & \\
\hline
\texttt{isGQ} & \texttt{isGG} & \texttt{isQQbar}\\ \texttt{isQQNeighbor} & \texttt{isQQOther} & \\
\hline
\texttt{numPartons} & \texttt{numPartonsPt20} &  \texttt{numPartonsPt100} \\
\texttt{numPartonsPt200} & \texttt{numB} & \texttt{numC} \\
\hline
 \texttt{firstkt} & \texttt{highestkt} & \texttt{numHighKt} 
 \\
\texttt{numHighKtMerge} & \texttt{hasPS} & \\
\hline
\end{tabular}}
\end{center}
\caption{Input variables used in the reweighting \dnnre}
\label{tab:reweightInputVars}
\end{table}

The variables used for these samples are shown in Table~\ref{tab:reweightInputVars}. For the incoming partons, five flag variables relating to flavors of the incoming partons are provided. These variables indicate whether the incoming partons are a gluon and a quark, two gluons, a quark and an antiquark, two quarks with a pdg id differing by one, or some other combination of quarks. These different classes correspond to different types of Feynman diagrams, so are important distinctions. Including the flavors tended to consistently bias the relative fraction of b, c, and light flavored quarks beyond the size of their systematic uncertainties without improving anything else, so they were not included in the final training. The observed bias likely came from the network learning a false correlation based on this variable. Additionally, the momentum distributions of incoming partons did not improve reweighting performance and were excluded from the final training inputs, as they were accurately predicted despite not being present during training.

To capture the information on the vector boson, \pt, $\eta$, and $\phi$ of the outgoing neutrinos are included, sorted by \pt. The mass is not included, as it is set to 0 for these light particles. Only the flavor of the original vector boson is included, as the flavor of the leptons was found to be unimportant. The reweighting function modeled lepton flavors accurately even when they were excluded. The vector boson is not included itself, as it was found that including its decay products and a flag dictating the flavor of the vector boson is sufficient.

The \pt, $\eta$, and $\phi$ values of the outgoing partons are included. Similar to the incoming partons, the exact flavors of the outgoing partons are not included, but the numbers of outgoing b and c quarks are included. The number of partons is included, along with the number above the merging scale (20\,GeV), above 100\,GeV, and above 200\,GeV. Finally, a flag for whether any of the $k_T$ values were below the merging scale, indicating the use of the parton shower, is included.

There are numerous variables based on the $k_T$ between the partons. The $k_T$ between two partons is defined as 
\begin{equation}
    k_T(p_1, p_2) = \min\{\pt(p_1),\pt(p_2)\}\Delta R(p_1, p_2),
\end{equation}
and in the case of a single parton, the $k_T$ is just the parton's \pt. 
The $k_T$ of all the outgoing quark combinations is included, along with the highest total $k_T$ and the number of pairs above the merging scale. The partons are clustered using the $k_T$ algorithm, and the $k_T$ of each successive merging is included. The first merging $k_T$ is included separately. There is a flag indicating that the parton shower was active if the $k_T$ of any of the pairs is less than the merging scale. A separate counter for the number of merges above the merging scale is also included. These variables are strongly correlated with the negative weights from the parton shower.

\subsection{\label{sec:training_variable_closure}Closure on training variables}

The testing and training set for the reweighting models use all Sherpa events with an event number equal to 1 modulo 100. Here, $40\%$ of these events are included in the test set, and $60\%$ are included in the training set. All validation plots are generated using the test set.

\begin{figure}[h]
    \centering
    \includegraphics[width=0.3\textwidth, valign=m]{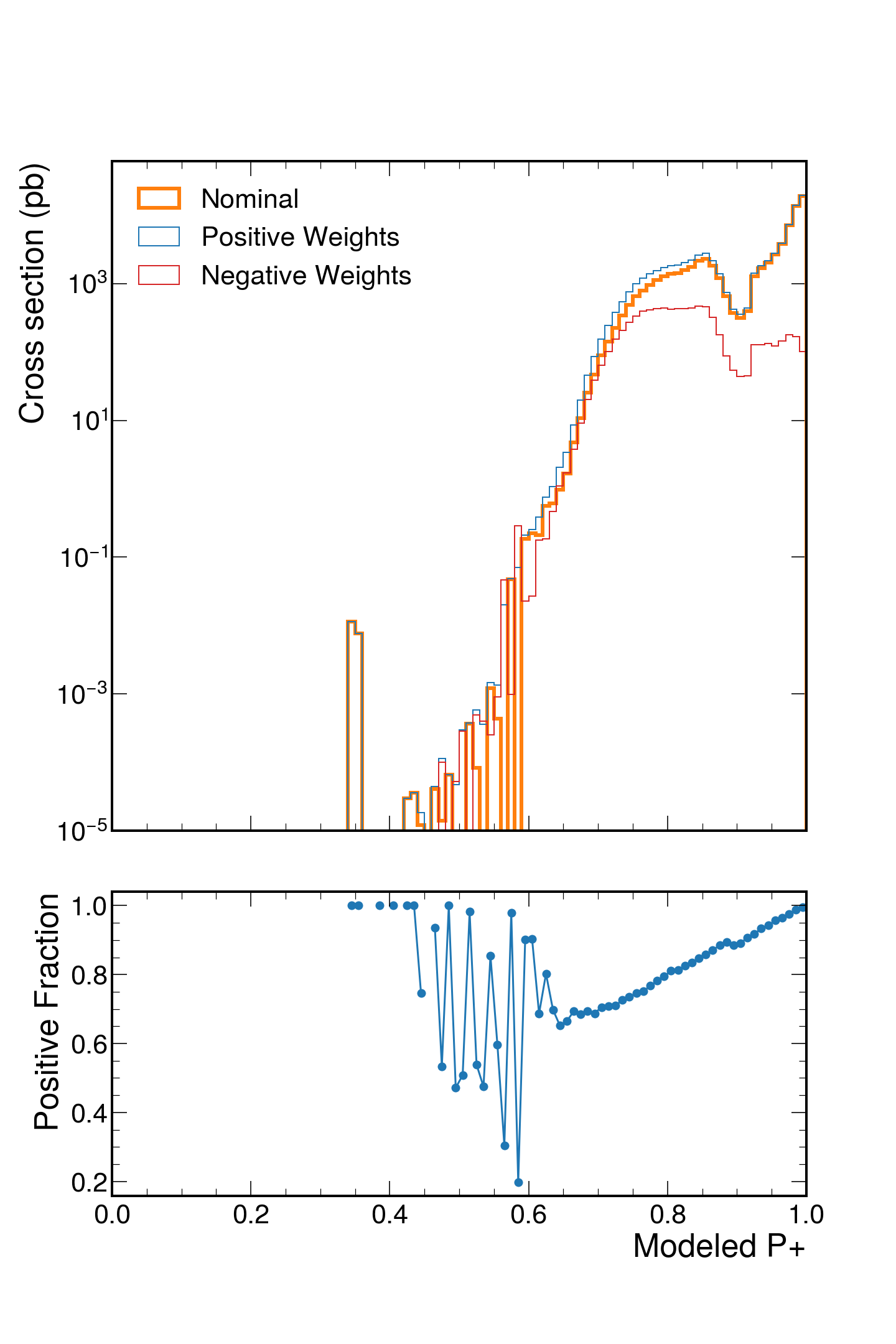}
    \includegraphics[width=0.3\textwidth, valign=m]{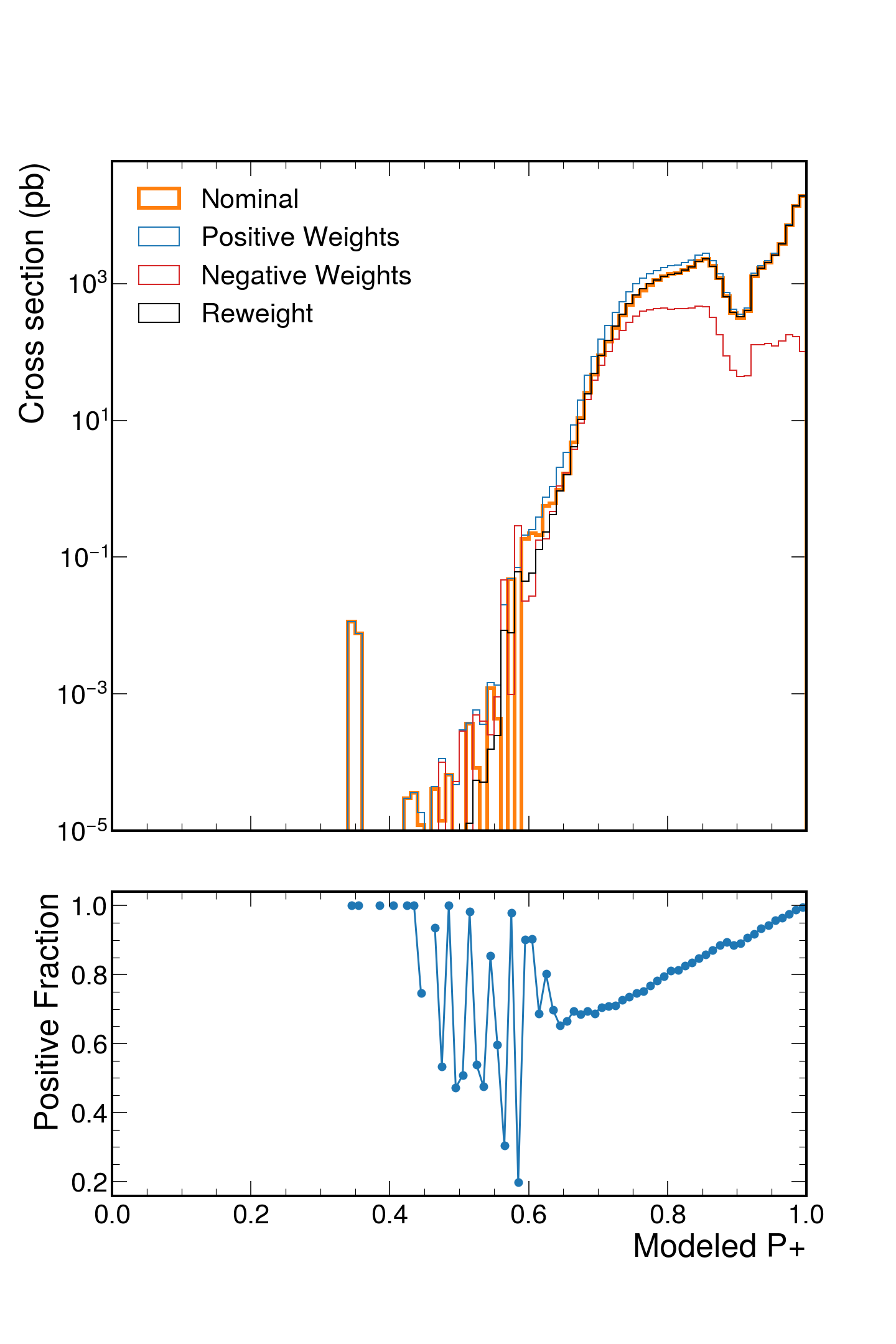}
    \caption{Histograms of the nominal $P_+$ distribution and the distribution of the positive and negative weights. The left plot has just the nominal distribution, while the right plot adds the reweighted distribution on top. These plots are normalized to the full cross sections of the processes.}
    \label{fig:reweight_dnn}
\end{figure}

\begin{figure}[h]
    \centering
    \includegraphics[width=0.3\textwidth, valign=m]{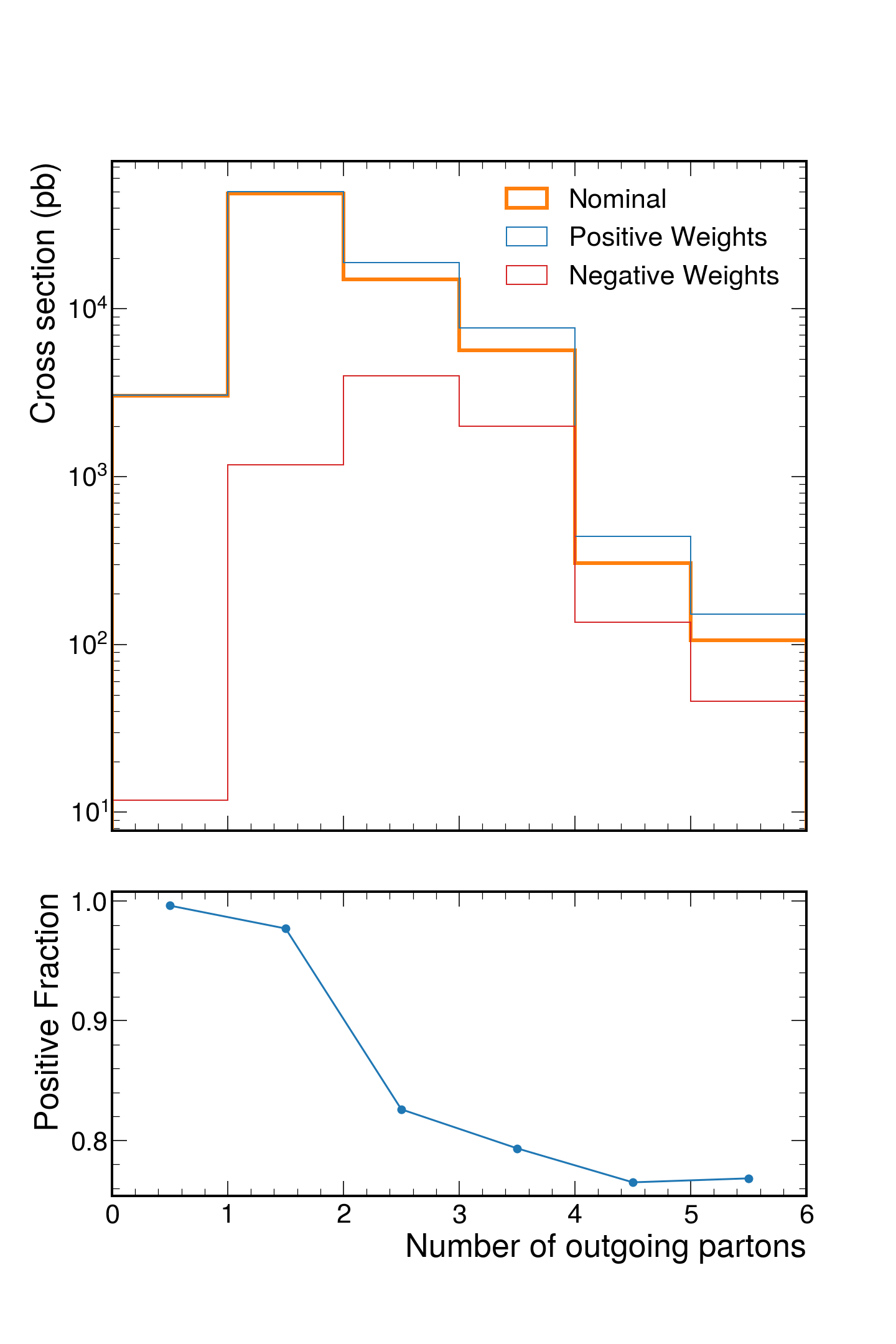}
    \includegraphics[width=0.3\textwidth, valign=m]{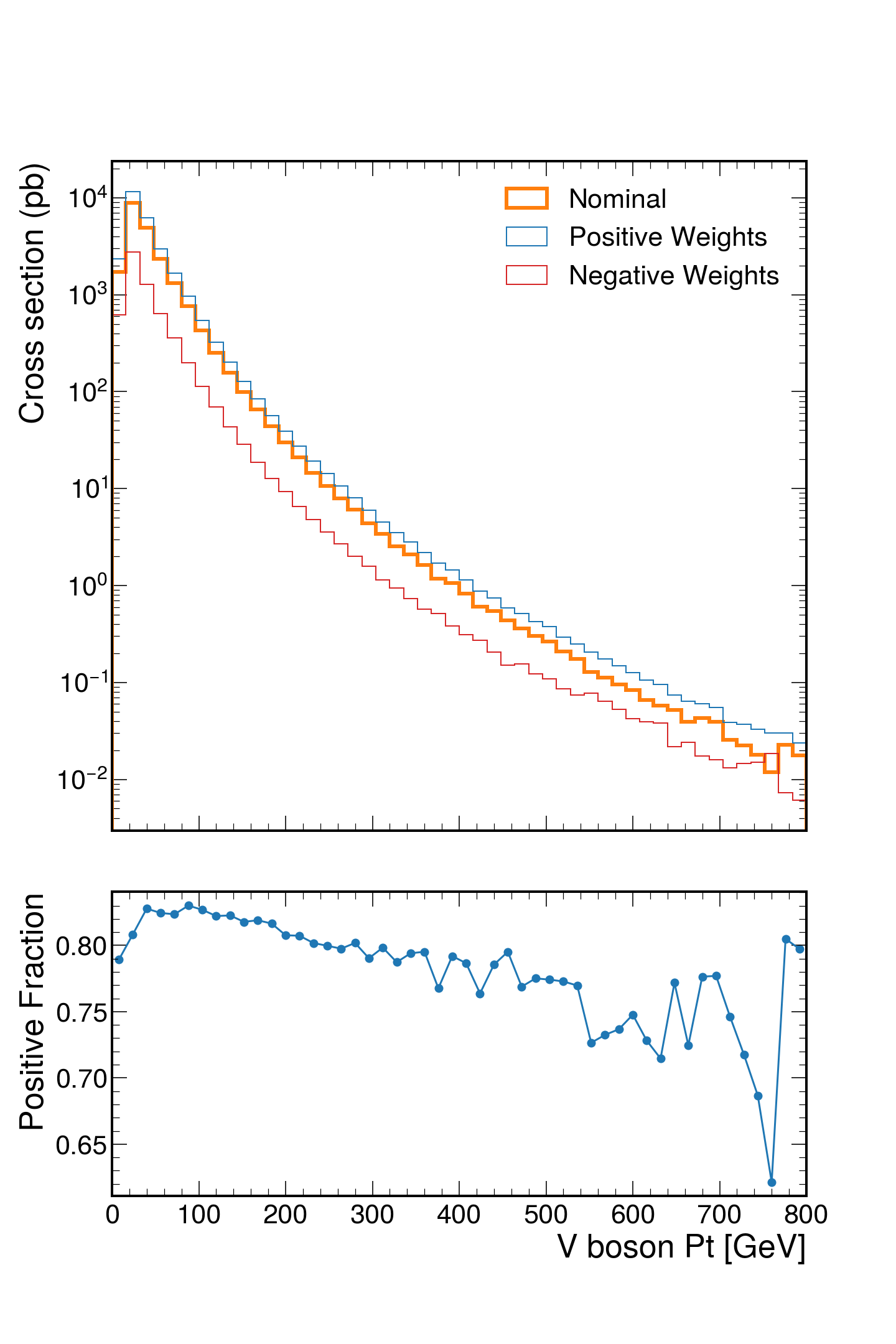}
    \caption{Histograms of the fraction of positive weight events. The left plot shows the fraction as a function of the number of partons. The right plot shows the fraction as a function of VpT when there are 2 or more partons.}
    \label{fig:frac_positive}
\end{figure}

The distribution of $P_+(\vec{x})$ modeled with the \dnnre is shown in Figure~\ref{fig:reweight_dnn}. The reweighting factor is $g(\vec{x}) = 2P_+(\vec{x}) - 1$ and is $\geq0$ if the modeled $P_+(\vec{x})\geq0.5$. As the output is almost entirely above 0.5, the reweighting is positive for nearly all $\vec{x}.$ 
The distribution falls into two main peaks: the lower output region corresponding to 2 or more outgoing partons and the higher output region corresponding to 0 or 1 parton. Plots of the fraction of positive weights as a function of number of partons and VpT can be seen in Figure~\ref{fig:frac_positive}.The fraction of positive weights increases linearly with the modeled $P_+(\vec{x})$ output as expected and without any significant bias. The reweighted distribution closes well with the original distributions. The only visible deviations come once the statistical precision of the original distribution becomes extremely poor and account for a tiny fraction of the total events.

\begin{figure}[h]
    \centering
    \includegraphics[width=0.3\textwidth, valign=m]{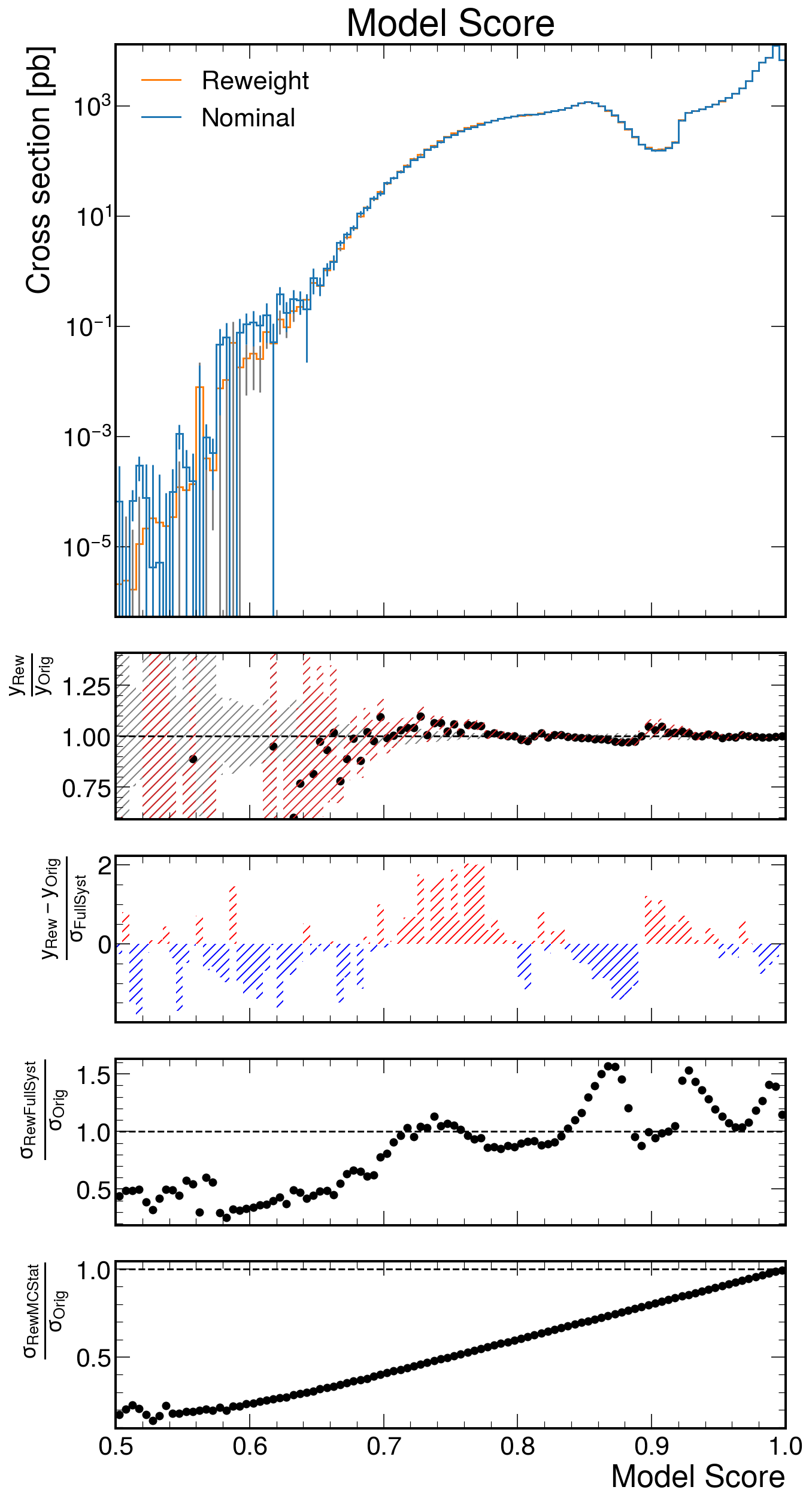}
    \includegraphics[width=0.3\textwidth, valign=m]{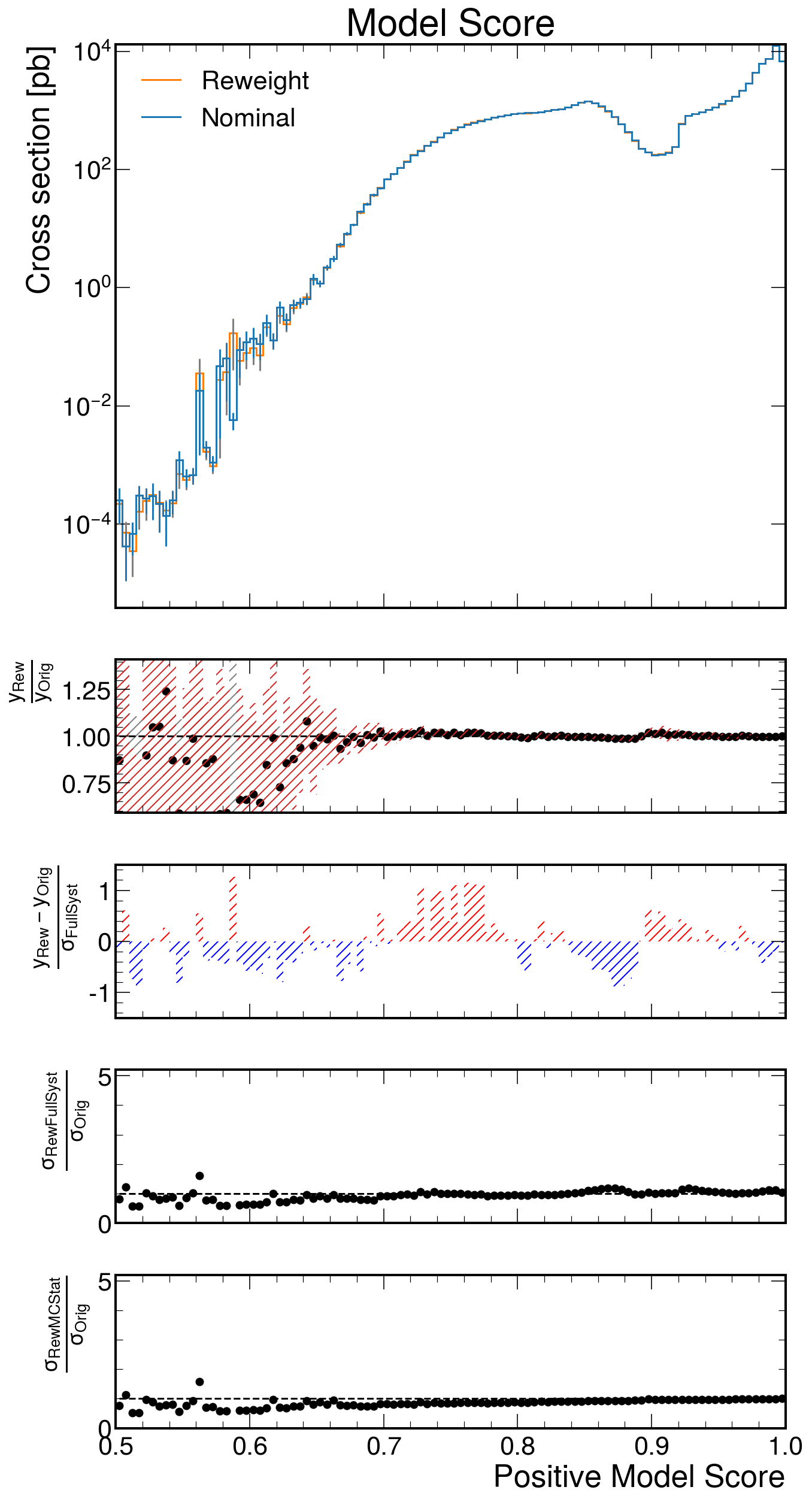}
    \includegraphics[width=0.3\textwidth, valign=m]{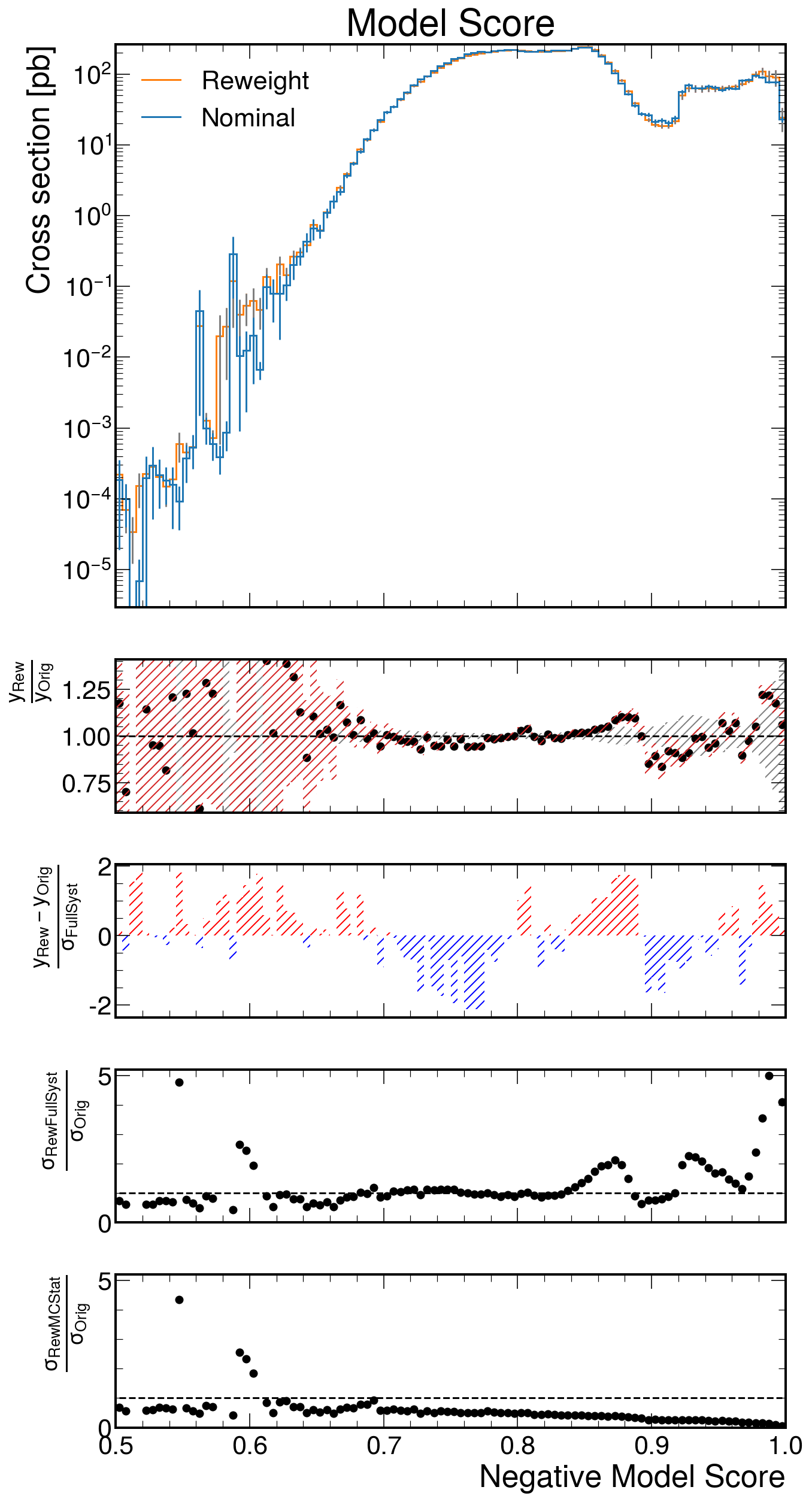}
    \caption{Closure on reweighting of the probability of being positive to the nominal distribution (left), the positive distribution (middle), and the negative distribution (right).}
    \label{fig:rw_closure}
\end{figure}

Further, the same plots as in Figure~\ref{fig:qm_closure} can be made to fully test the closure by reweighting to the positive and negative distributions. The center values are set to the average bin value from each of the 20 alternative histograms. The systematic uncertainties are the sum of all 20 PCA directions in quadrature from the 20 alternative $g(\vec{x})$ fits. The systematic variations vary the weights for the events, but the probability of positive is fixed to its average. This prevents the simultaneous variation of two variables, allowing for a consistent visualization. These are in Figure~\ref{fig:rw_closure}. The closure is generally good, and the systematic uncertainties can account for the observed discrepancies. The reduction in MC statistical uncertainty is visible in the bottom panels. The second panel from the bottom show the change in net uncertainty when including the systematic. As the systematic uncertainties are correlated across all the bins, their treatment here as binwise is an approximation. The regions where the net uncertainty is not lowered by the reweighting generally combine a non-negligible uncertainty with a large number of events. This is expected, as if there are a large number of events the MC stat uncertainty will be negligible, even with a large fraction of negative weights. As will be seen later, the main benefit of this method is where there are significantly fewer events in each histogram bin.

\begin{figure}[h]
    \centering
    \includegraphics[width=0.2\textwidth, valign=m]{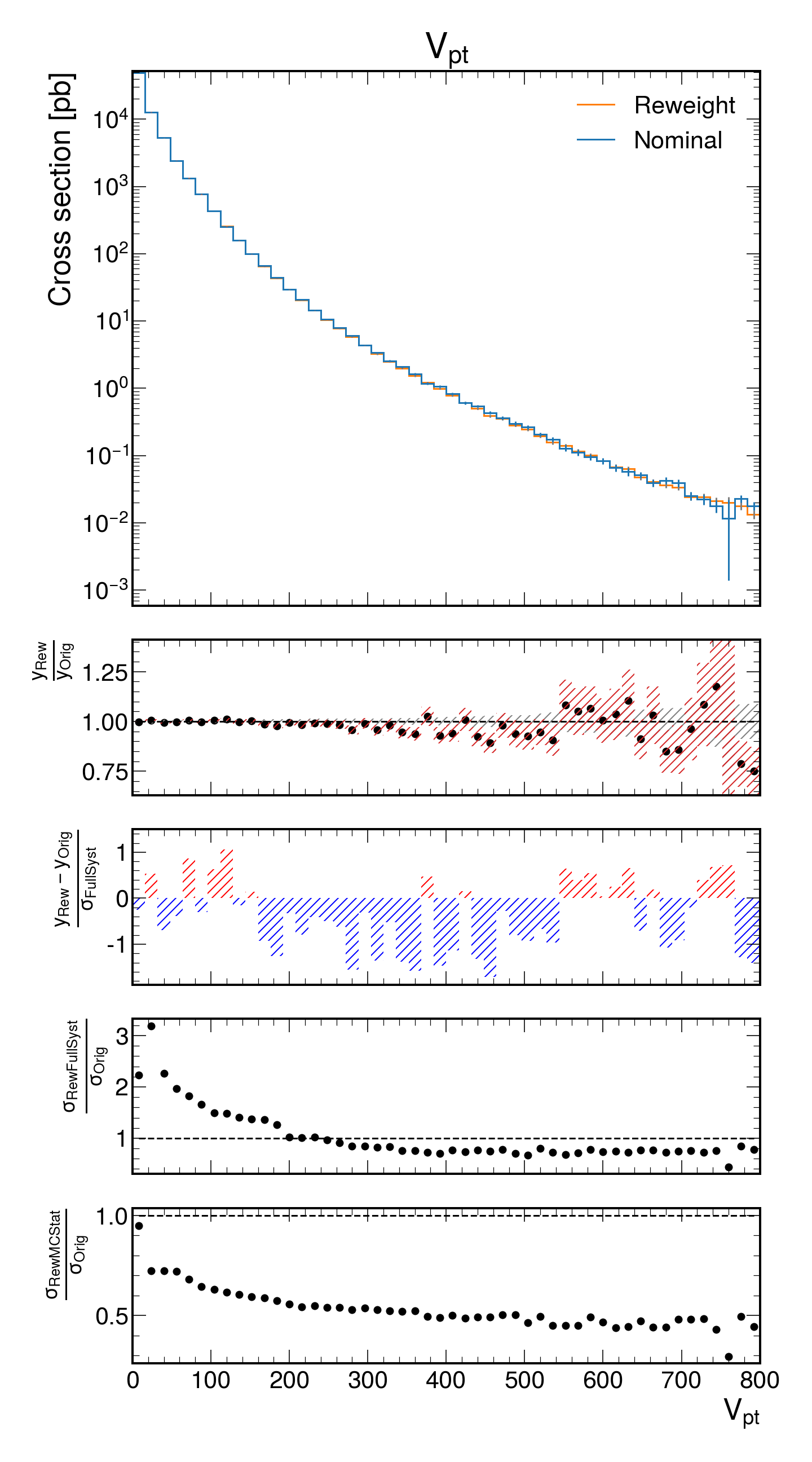}
    \includegraphics[width=0.2\textwidth, valign=m]{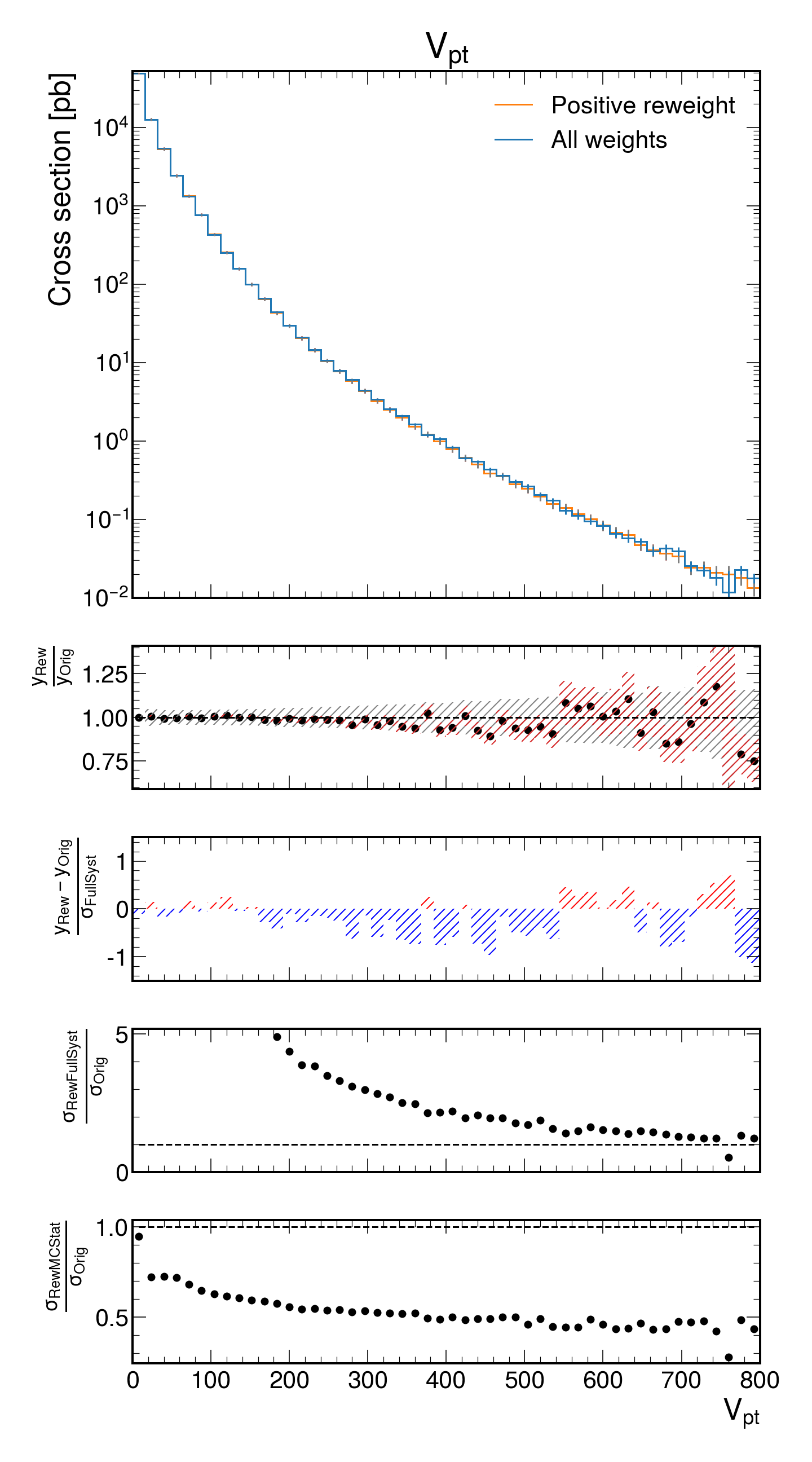} \\
    \includegraphics[width=0.2\textwidth, valign=m]{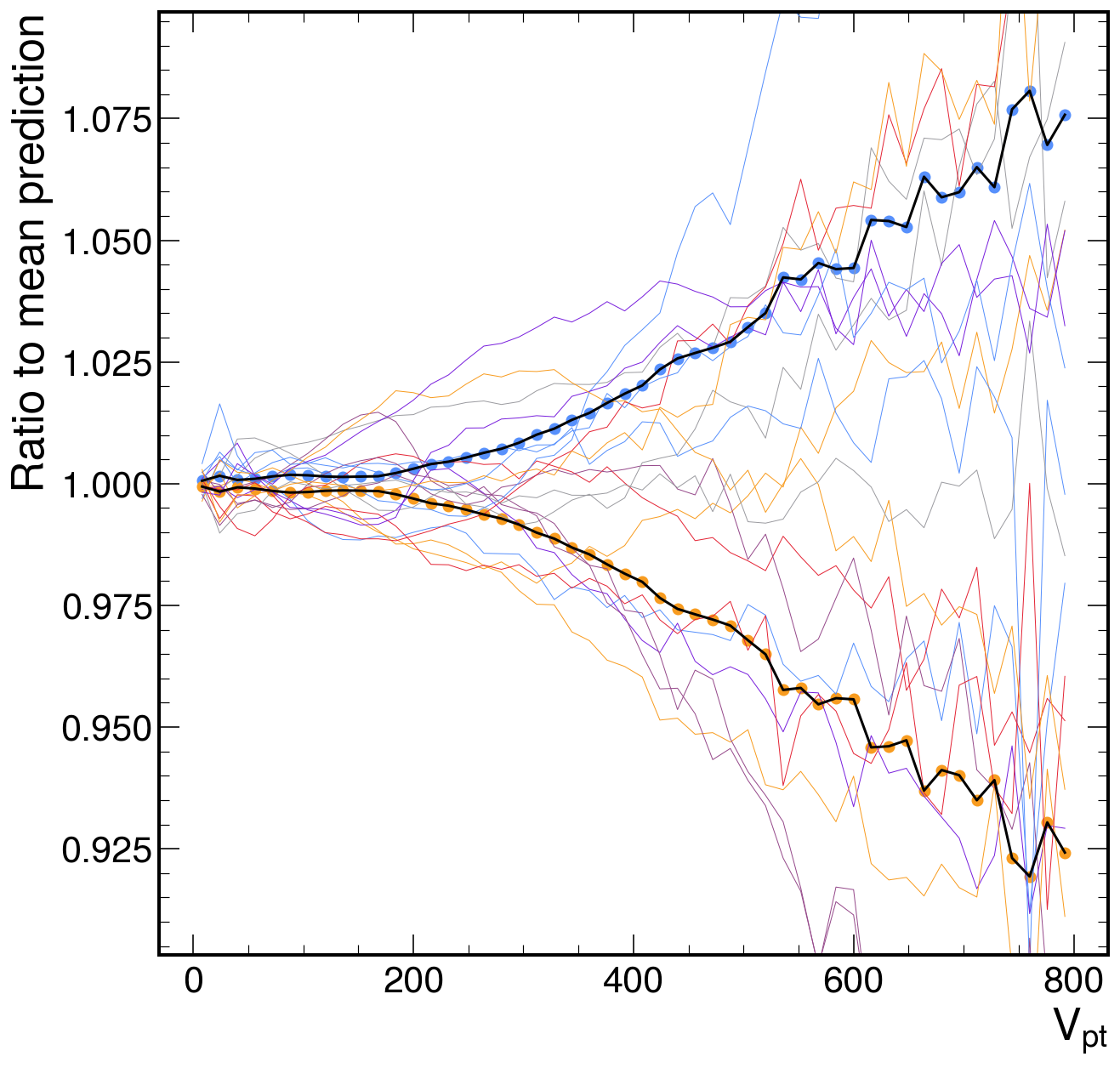}\\
    \caption{Plots of the generator level closure of the reweighting. The top panels show the reweighted and nominal distributions with their error bars. The ratio panel shows the ratio of the reweighted distribution to the nominal one and the corresponding uncertainty from the weights in red. The uncertainty from the systematic is shown in gray, while the statistical uncertainty is in red. The third panel shows the pull on the combined uncertainties of the two plots, plus the systematic, on the deviation between the reweighted and nominal distributions. The fourth panel shows the ratio of the net uncertainties between the reweighted histogram and the nominal histogram. The systematic uncertainty is included in the value for the reweighted histogram. The final panel shows the ratios of just the MC statistical uncertainties. The left plot uses the PCA-based systematic and the right plot uses the event-based systematic. The bottom plot shows the first PCA vectors, along with the 20 shape variations. } 
    \label{fig:gen_validation}
\end{figure}

With the reweighting function defined, the closure on one of the input variables is checked. The $\ptV$ is shown Figure~\ref{fig:gen_validation}. The left plot shows the PCA-based systematic, and the right plot shows the event-based systematic. The center values are set to the average bin value from each of the 20 alternative histograms. The shaded band represents the systematic uncertainties, which are the sum in quadrature of all 20 PCA directions from the 20 alternative $g(\vec{x})$ fits for the PCA uncertainty and the up and down variations for the event-based systematic. For these variables, the first PCA vector is a correlated pull up or pull down, the effect of which increases with \pt. The bottom panels show that the statistical uncertainty is reduced by up to a factor of 2, but the second panel from the bottom shows that when including the systematic the net uncertainty can increase. It is evident that the PCA-based systematic is much smaller that the event-based one, and the regions where it still increases the net uncertainty are those with very large numbers of events.

\subsection{\label{sec:signal_region_definition}Signal region definition}
As the reweighting closes well for the generator variables, the next step is to construct and check a signal-versus-background \dnnsvb in the signal-enriched region. The \dnnsvb model is trained using the original weighting, but the reweighted samples could also be used for the training. The signal region is defined by the selections listed in Table~\ref{tab:selection_criteria}. The events used for training the reweighting \dnnre are excluded as well.

\begin{table}[h!]
\begin{center}
\resizebox*{0.6\textwidth}{!}{
\begin{tabular}{|l|l|}
\hline
Selection criterion & Requirement \\
\hline
missing energy & $> 180$ GeV \\
\hline
leptons & 0 good leptons \\
\hline
jets & 2$-$3 good jets \\
\hline
dijet $p_T$ & $> 150$ GeV \\
\hline
$m_{jj}$ & $ = 90-150$ GeV \\
\hline
B-tagging & btag $ > 0.5$ \\
\hline
j$_1$ $p_T$ & $>$ 60 GeV \\
\hline
j$_2$  $p_T$ & $>$ 35 GeV \\
\hline
MET-dijet $\Delta\phi$ & $ > 2.2$ \\
\hline
MET-dijet $p_T$ ratio & $0.5 < \frac{MET}{p_T} < 2$ \\
\hline
\end{tabular}}
\caption{Event selection criteria}
\label{tab:selection_criteria}
\end{center}
\end{table}

The \dnnsvb is trained on half the events passing these selections and evaluated on the other half. The variables used for training are in Table~\ref{tab:dnn_sr_variables}. They describe most of the relevant kinematics to the system. Again, the original weights are used for the \vjets samples, though the reweighted ones could be used as well. The histogram of the \dnnsvb score is binned such that the signal distribution is flat, while the MC statistical uncertainty of the background in each bin is less than $25\%$. Given this approach, lower MC statistical uncertainties allows for more bins and greater sensitivity, in this case granting three times as many bins and up to a $50\%$ increase in Asimov significance.

\begin{table}[h!]
\begin{center}
\resizebox*{0.9\textwidth}{!}{
\begin{tabular}{|l|l|l|l|l|}
\hline
$\pt(\textnormal{V})$  & $\phi(\textnormal{V})$ & nJets & - & - \\
\hline
$\pt(j_1)$  & $\eta(j_1)$ &  $\phi(j_1)$ &  m$(j_1)$ &  btag$(j_1)$    \\
$\pt(j_2)$  & $\eta(j_2)$ &  $\phi(j_2)$ &  m$(j_2)$ &  btag$(j_2)$    \\
\hline
$\pt(H)$  & $\eta(H)$ &  $\phi(H)$ &  m$(H)$ & - \\
\hline
$\Delta\eta(jj)$ & $\Delta\phi(jj)$ & $\Delta R(jj)$ & - & - \\
$\Delta\phi(\textnormal{MET}, j_1)$ & $\Delta\phi(\textnormal{MET}, j_2)$ & $\Delta\phi(\textnormal{MET}, jj)$ & - & - \\
$\Delta\phi(\textnormal{V}, H)$ & $\pt(\textnormal{V})/\pt(H)$ & - & - & - \\
\hline
\end{tabular}}
\end{center}
\caption{Variables used for SR \dnnsvb training.}
\label{tab:dnn_sr_variables}
\end{table}

\subsection{\label{sec:PCA_systematics}PCA systematic variations}

\begin{figure}[h]
    \centering
    \includegraphics[width=0.4\textwidth, valign=m]{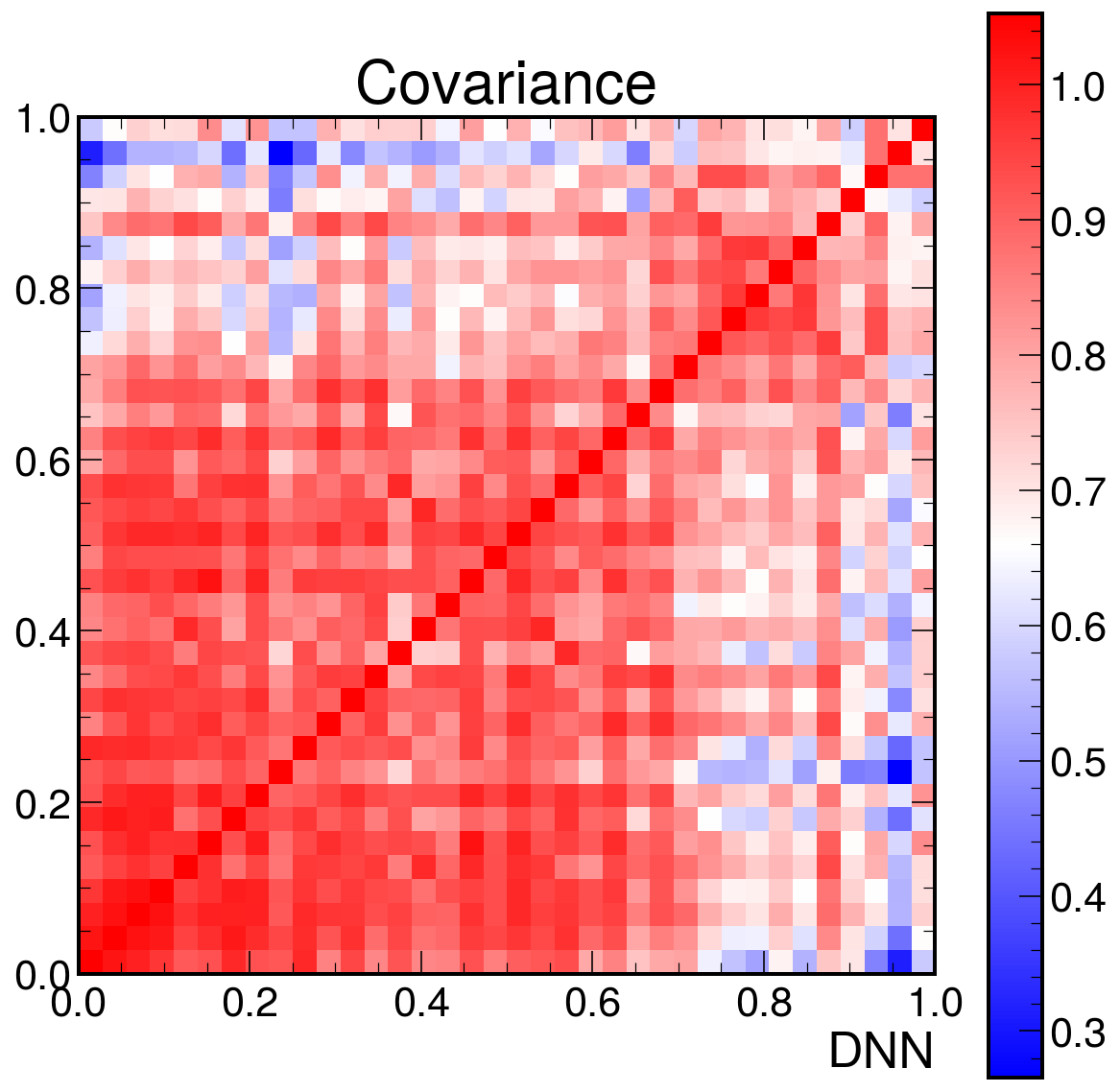}
    \caption{Covariance between the \dnnsvb histogram bins coming from the 20 alternative shapes. The different shapes are not normalized relative to each other, so there is a net up and down uncertainty.}
    \label{fig:sr_covariance}
\end{figure}

\begin{figure}[h]
    \centering
    \includegraphics[width=0.2\textwidth, valign=m]{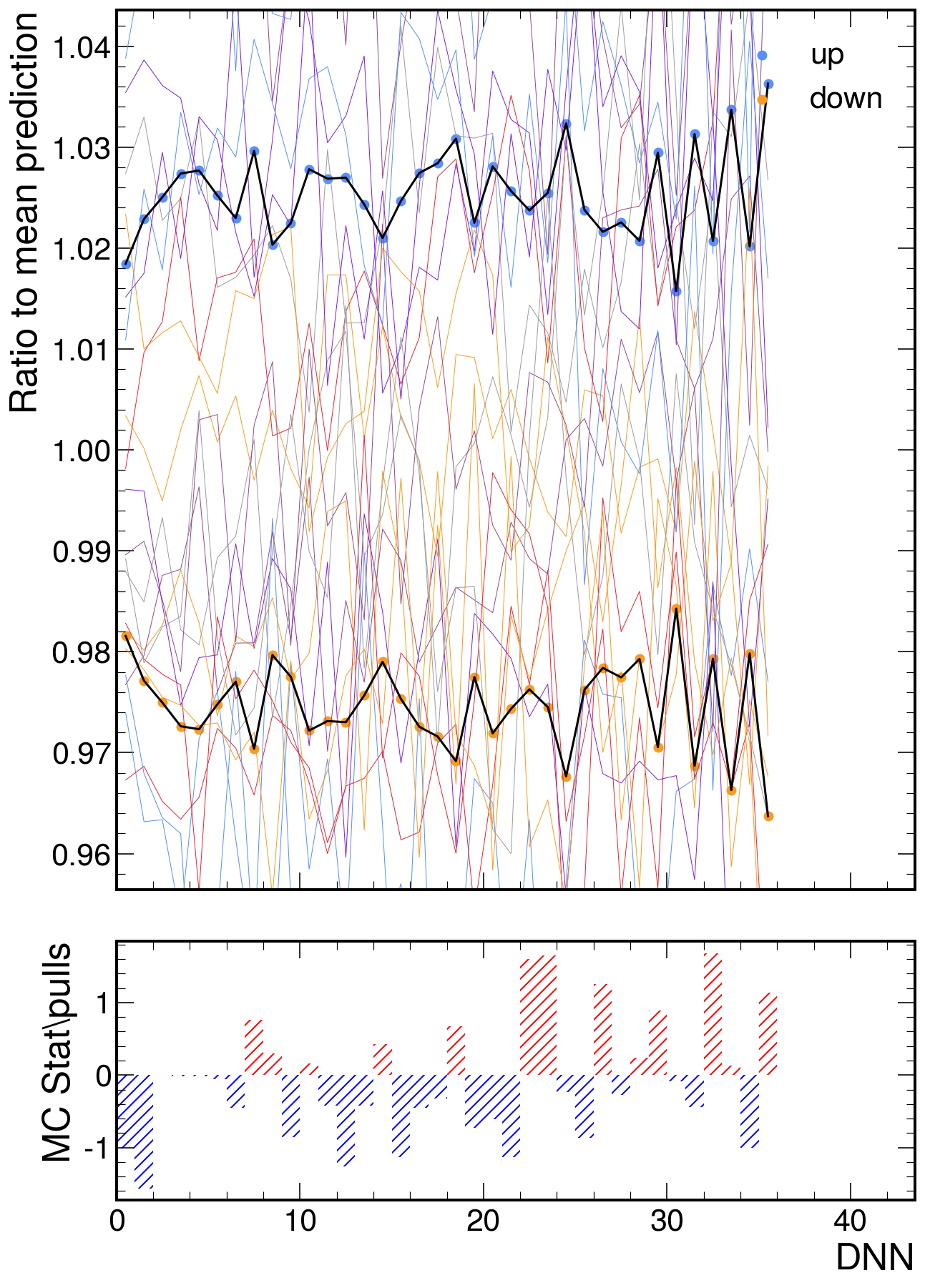}
    \includegraphics[width=0.2\textwidth, valign=m]{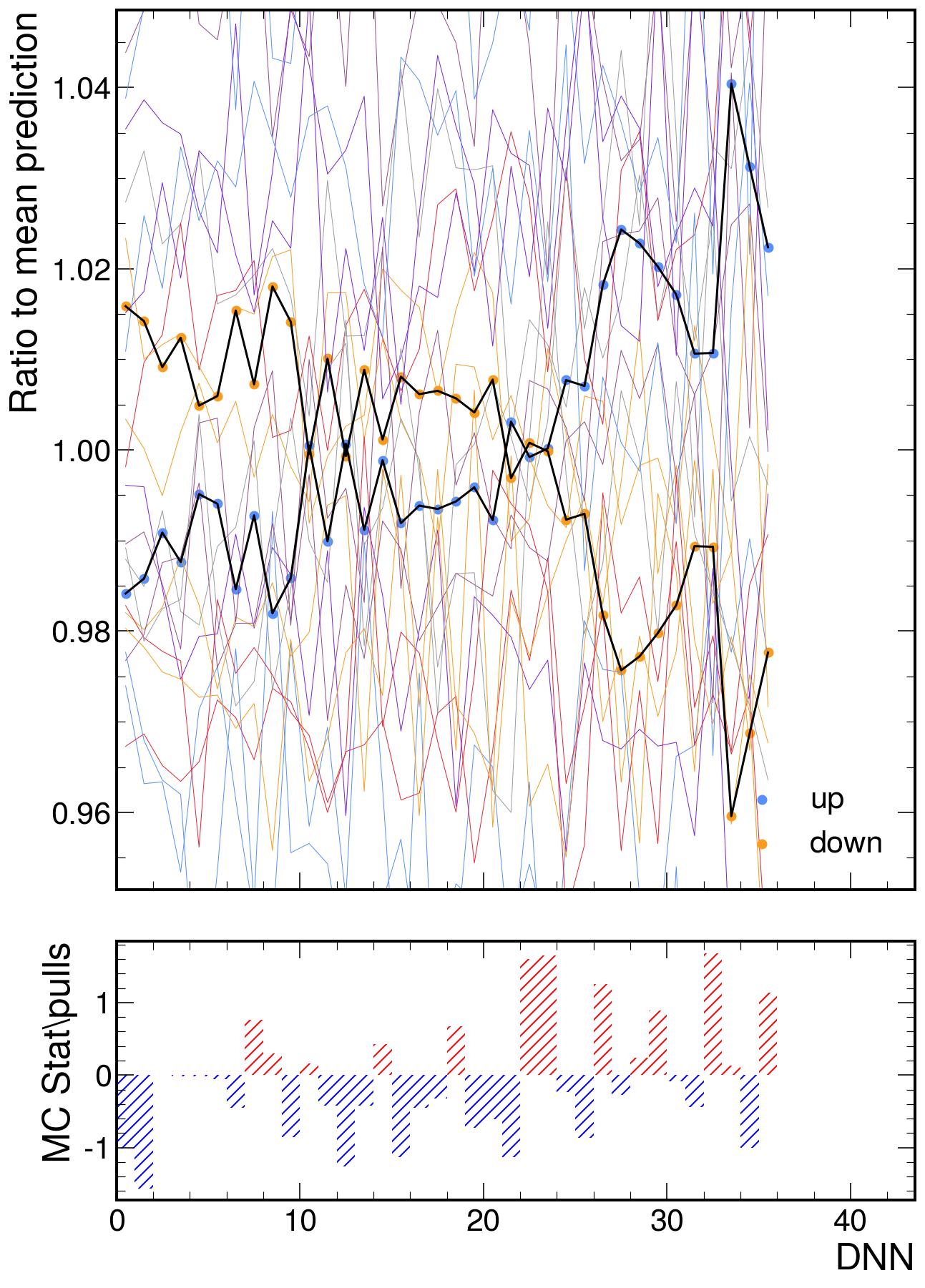}
    %\\
    %\includegraphics[width=0.2\textwidth, valign=m]{DNN_direction3.png}
    %\includegraphics[width=0.2\textwidth, valign=m]{DNN_direction4.png}
    \caption{The first two PCA vectors from the unnormalized histograms. The first is primarily a normalization shift and the second is an anticorrelated shift between low and high score bins. Subsequent vectors have smaller, more complex correlated shifts. The bottom panel in each plot shows the impact of including the systematic uncertainty on the pulls.}
    \label{fig:sr_pca_directions}
\end{figure}

With the signal region \dnnsvb trained, its PCA systematic uncertainties on it can be examined in detail. The full covariance matrix for the signal region bins is shown in Figure~\ref{fig:sr_covariance}, and the first two PCA vectors are in Figure~\ref{fig:sr_pca_directions}. The first PCA vector is an overall normalization uncertainty of $3\%$. The second is mostly an anticorrelated shift of the low and high \dnnsvb output regions. The other vectors are more complex and smaller. The first two are required to cover the anticorrelation between the low and high score bins. As the uncertainties are treated as binwise in the plots, the systematic is calculated by taking the sum of all 20 directions in quadrature.

\begin{figure}[h]
    \centering
    \includegraphics[width=0.2\textwidth, valign=m]{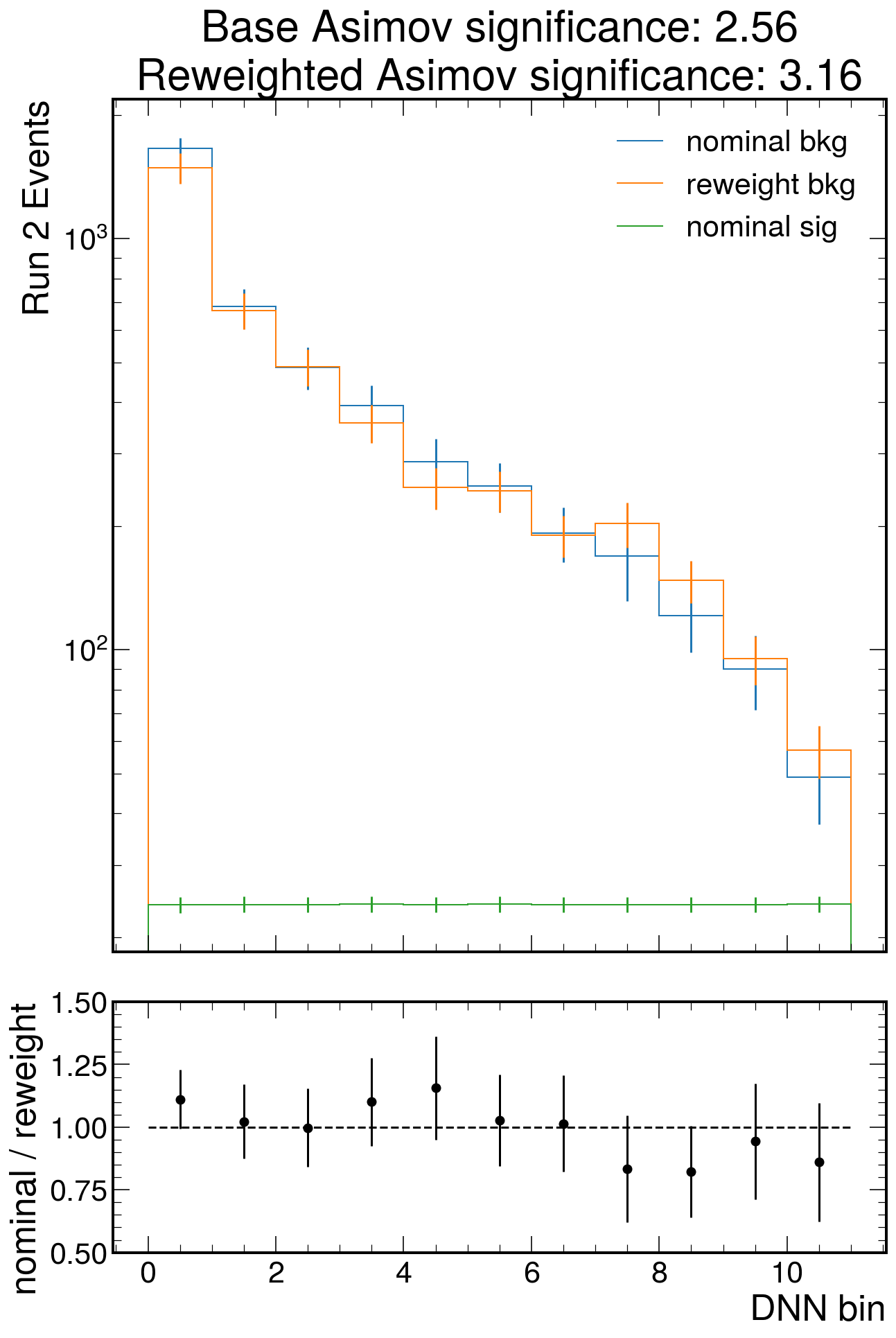}
    \includegraphics[width=0.2\textwidth, valign=m]{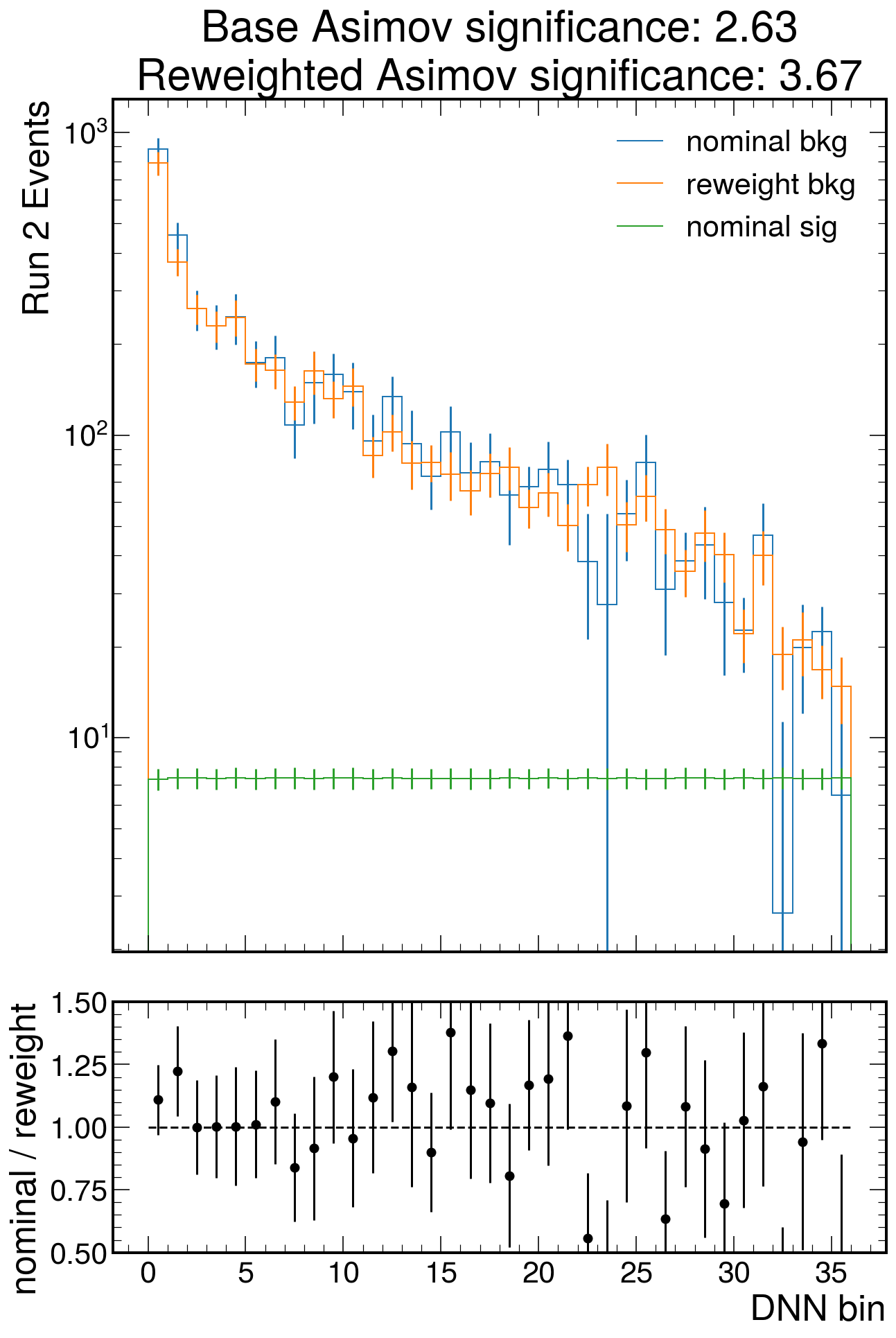}
    \includegraphics[width=0.2\textwidth, valign=m]{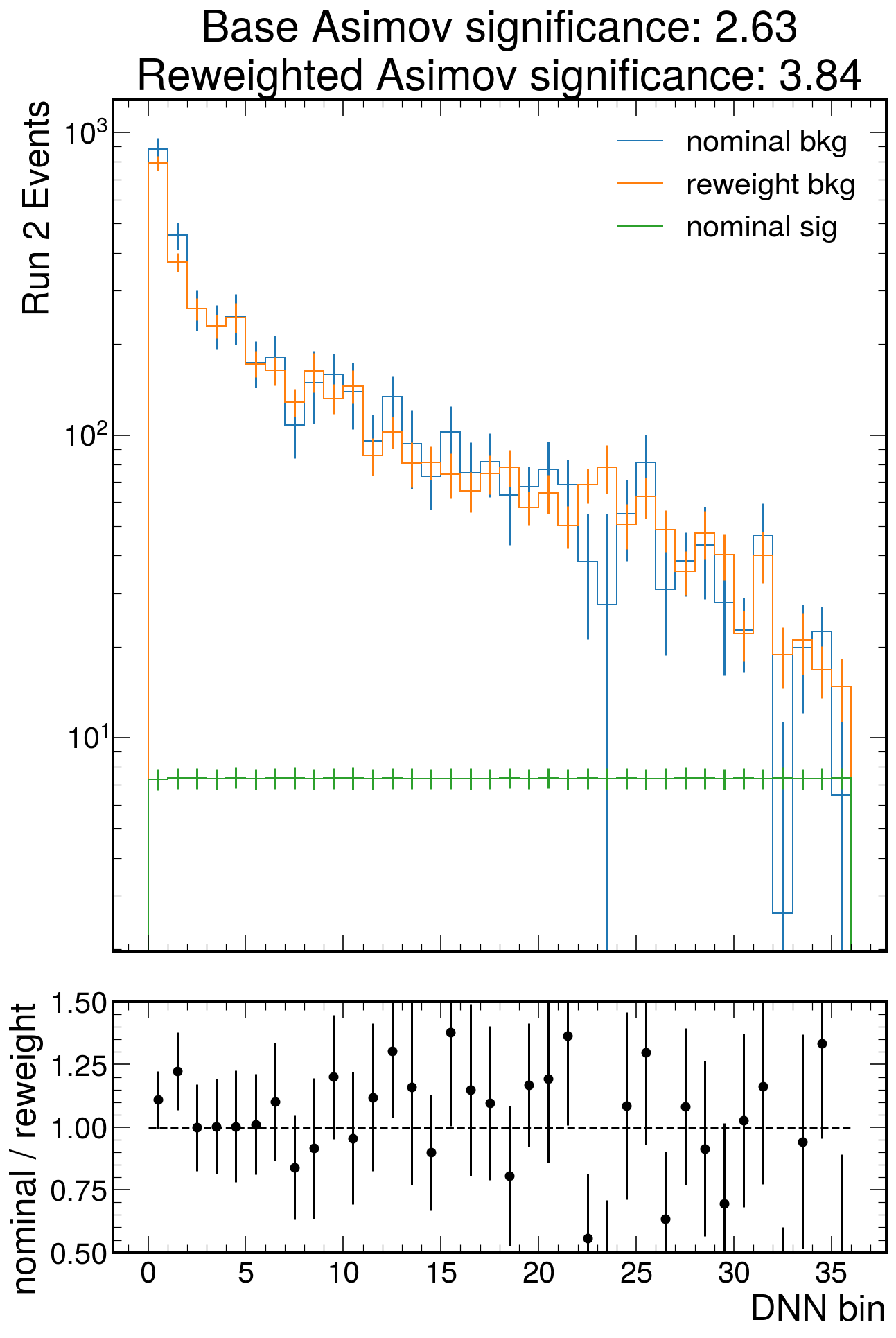}\\
    \includegraphics[width=0.2\textwidth, valign=m]{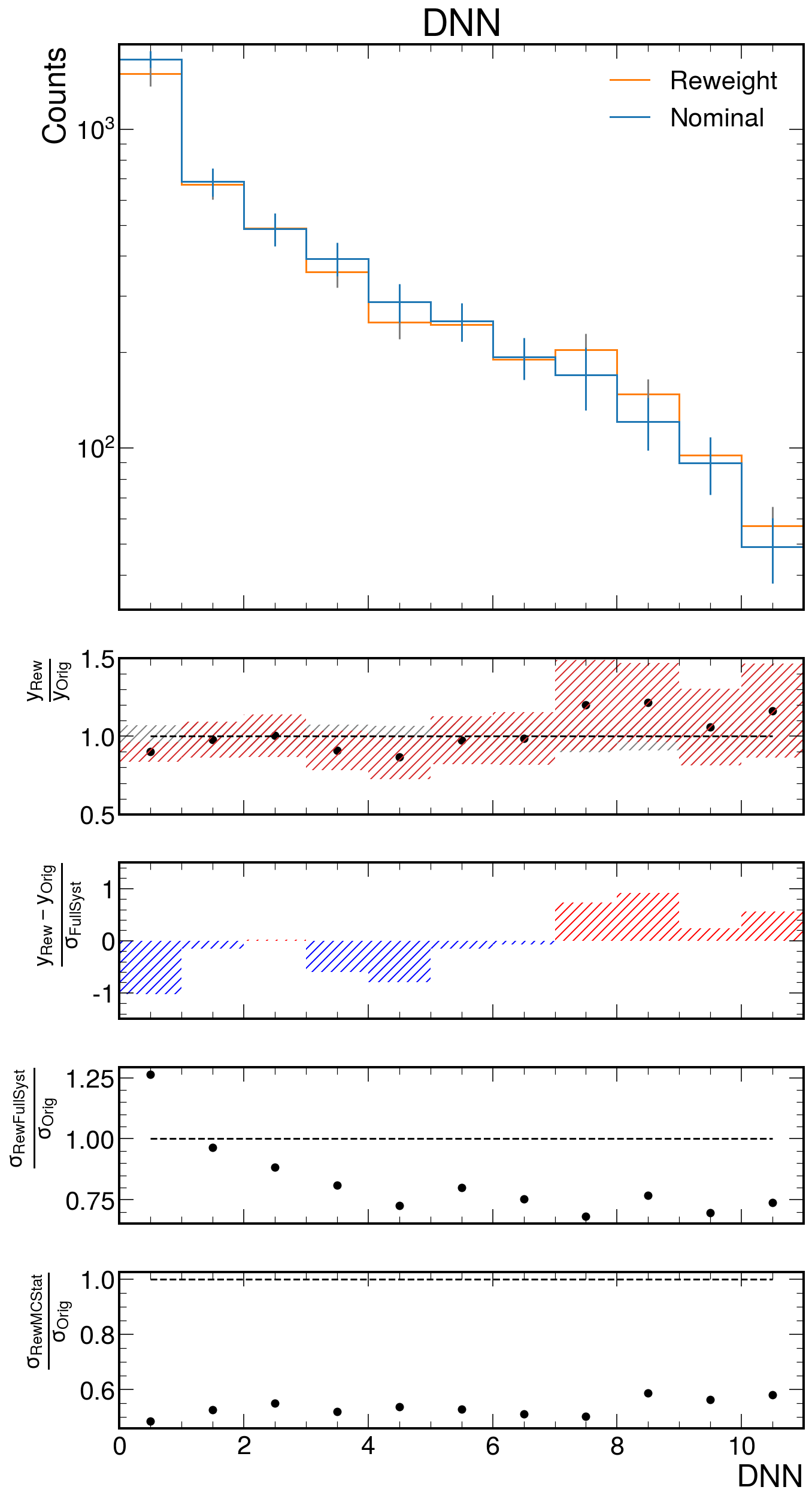}
    \includegraphics[width=0.2\textwidth, valign=m]{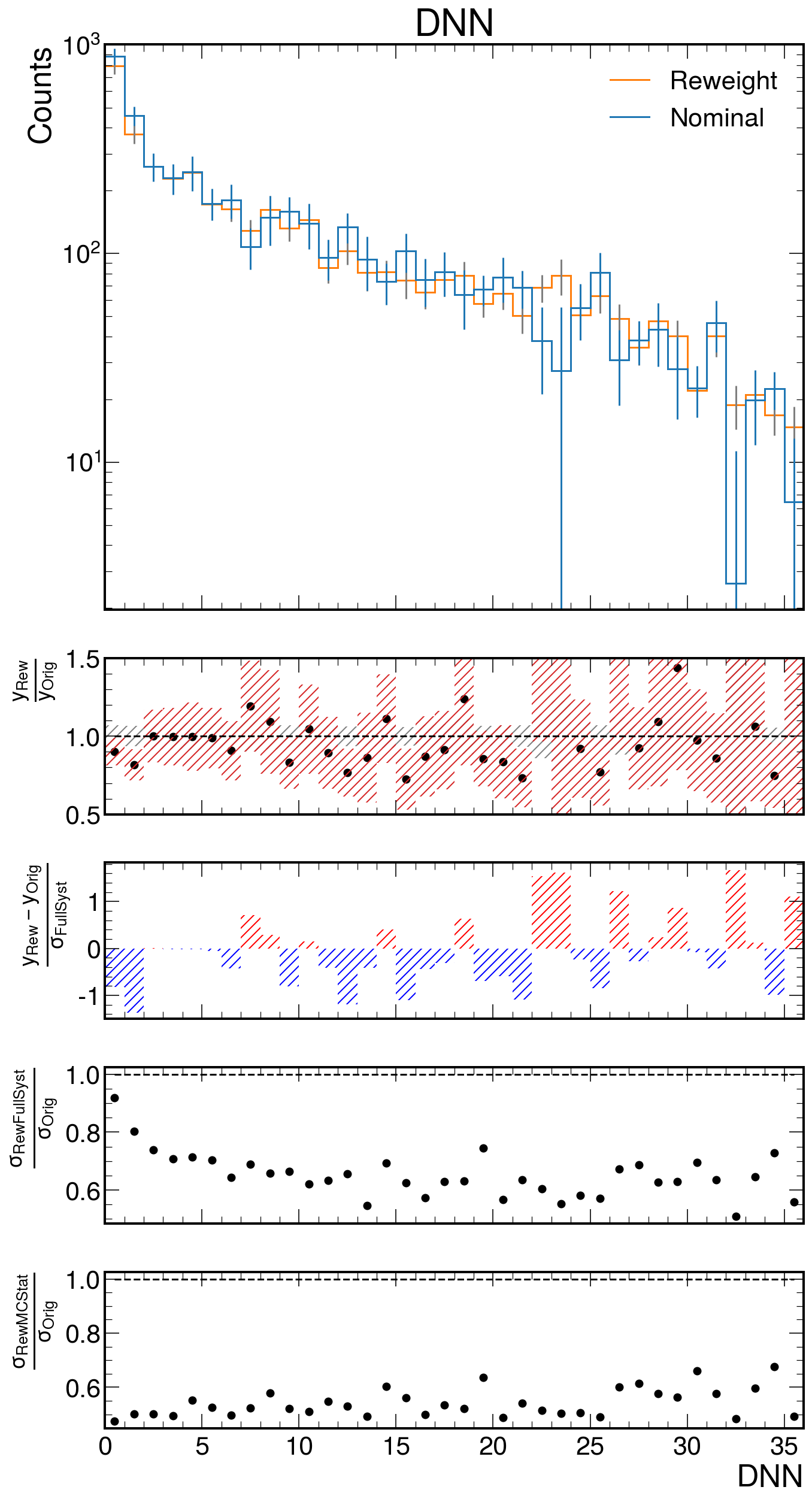}
    \includegraphics[width=0.2\textwidth, valign=m]{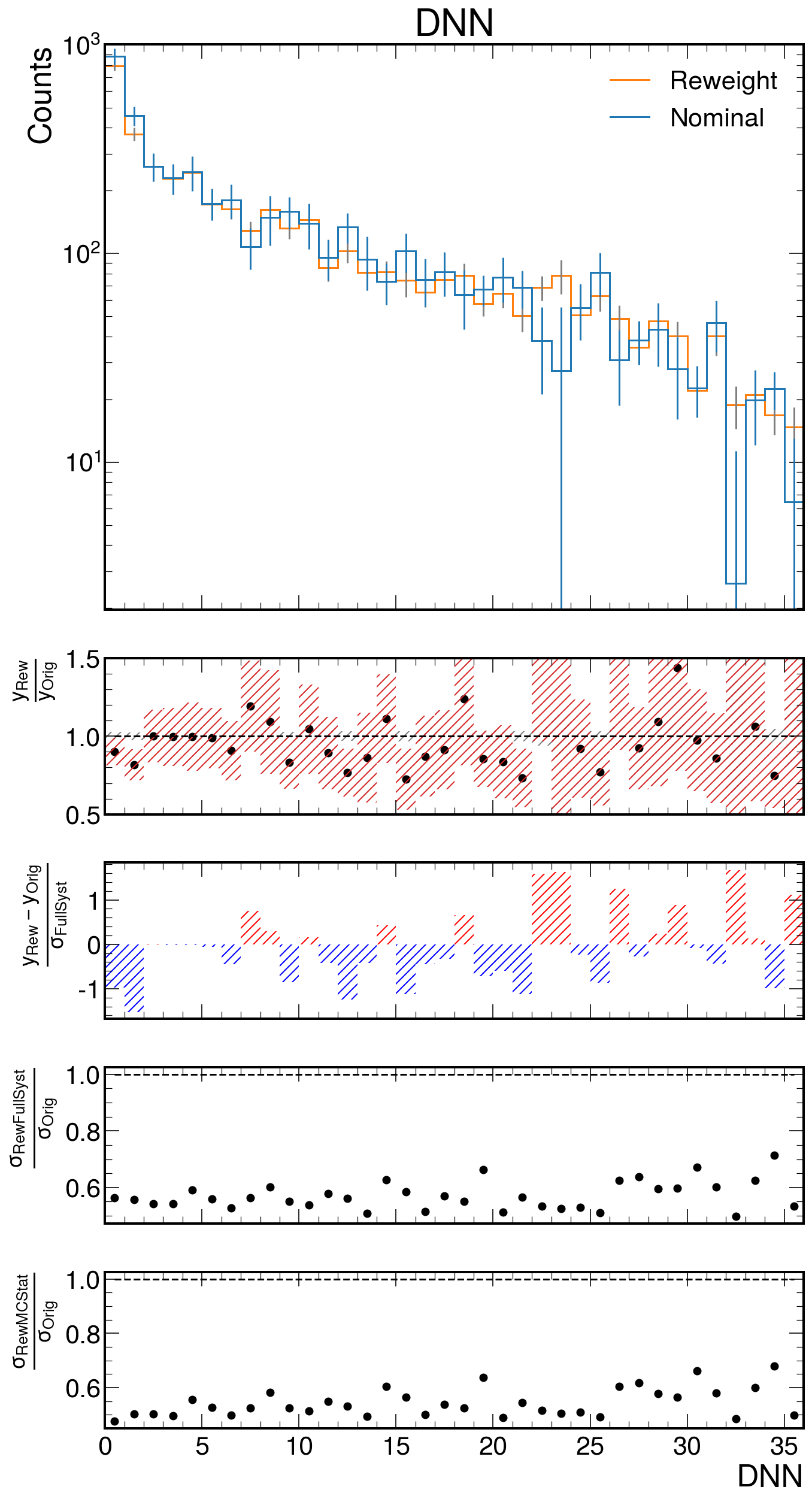}
    \caption{Plots of the SR region histograms. The leftmost plots correspond to using the nominal weights to determine the binning with a target 25\% maximum binwise uncertainty. The middle plots correspond to using a fully correlated bin-by-bin uncertainty with a 25\% target. The rightmost plots correspond to using the sum in quadrature of all PCA directions with a 25\% uncertainty. The expected Asimov significance obtained from using the nominal or reweighted samples is reported for all three options. } 
    \label{fig:sr_asimov}
\end{figure}

\begin{table}[h!]
\begin{center}
\resizebox*{0.8\textwidth}{!}{
\begin{tabular}{|l|l|l|l|}
\hline
Sample & Binning source & Systematic source & Asimov significance \\
\hline
Nominal & Nominal & - & 2.56 \\
Reweighted & Nominal & Event-based & 3.16 \\
Reweighted & Reweighted & Event-based & 3.67 \\
Reweighted & Reweighted & PCA & 3.84 \\
\hline

\end{tabular}}
\caption{Asimov Significance of the signal region using various strategies. The Sample column refers to whether the significance was calculated with the nominal or reweighted samples. The Binning source column refers to whether the bins were determined using the nominal or reweighted samples. The systematic source refers to the reweighting uncertainty used on the reweighted samples, if they are used. Finally, the Asimov significance is the expected sensitivity to the signal process given the binning and background.}
\label{tab:asimov_sig}
\end{center}
\end{table}

The resulting histograms with their uncertainties and Asimov significance \cite{Elwood_2020} are shown in Figure~\ref{fig:sr_asimov} and Table~\ref{tab:asimov_sig}. Using the rebinned samples with the PCA systematic gives an increase in sensitivity of almost $50\%$ compared to just using the nominal samples. The two distributions agree very well. Finer binning shows the reweighted distribution is less noisy than the original. 

The first column of histograms corresponds to using the original weights and has the smallest number of bins. The systematic uncertainty assigned to the reweighted histogram is the event-based value. Even without increasing the number of bins there is a substantial gain in expected significance due to the lower bin-by-bin uncertainties.

The second column assumes the scenario of full correlation between bins and shows a significant gain in the number of bins compared to using the original samples and a half sigma increase in significance. The bin-by-bin systematic uncertainty is combined with the base MC statistical uncertainty when calculating the binning. Due to the small number of events in each bin, it still provides a very small uncertainty in most of the bins, as in Eq.~\ref{eqn:uncertGain}. It is visible, however, that as the number of events climbs to about one thousand that the systematic uncertainty will become large enough to cancel out the reweighting gains. Thus, in any bin of significant size, the uncertainty would be much higher.

For the final column, the sum of all the PCA shape systematic uncertainties in quadrature is used. This yields a result which is nearly identical to the previous column, but with slightly reduced uncertainties and a slightly increased significance. In particular, the full systematic ratio stays at about $60\%$ for all the bins when using the PCA systematic uncertainties, while it increases to  $95\%$ in the first bin when using the event-based systematic. Thus, the PCA-based systematic provides consistent, expected behavior, while the event-based systematic diverges in the more populated bins.

\begin{figure}[h]
    \centering
    \includegraphics[width=0.2\textwidth, valign=m]{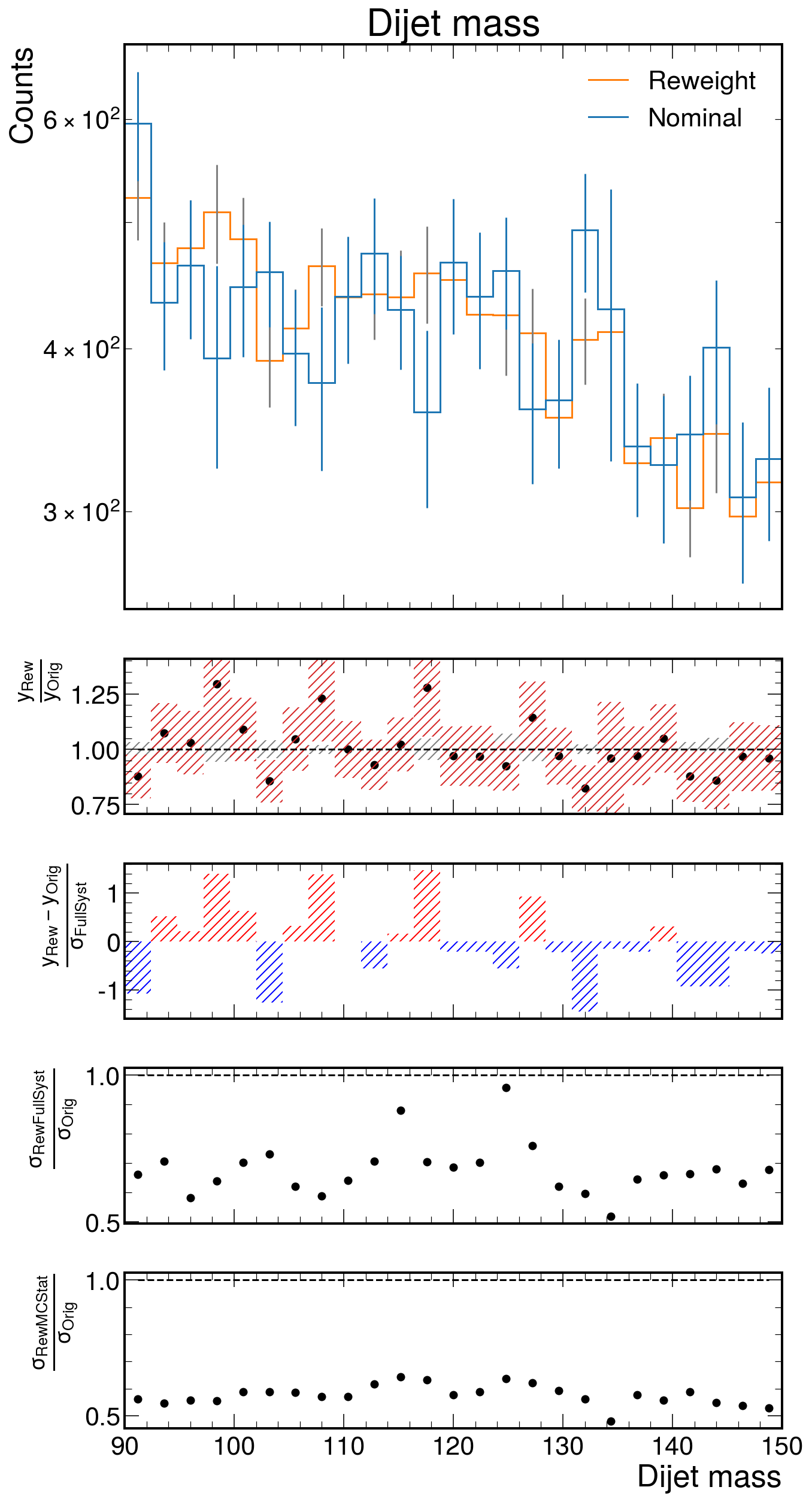}
    \includegraphics[width=0.2\textwidth, valign=m]{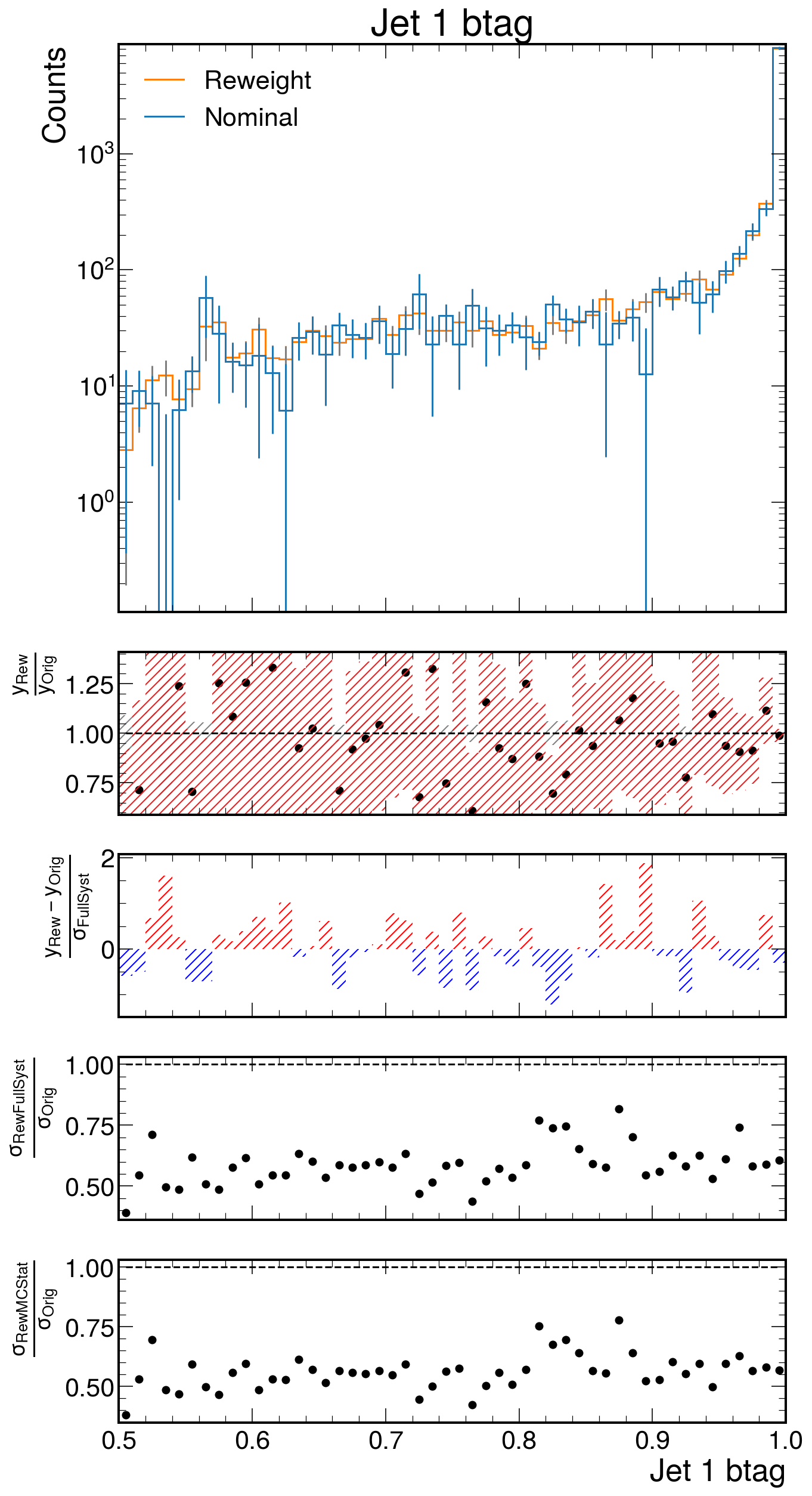}
    \includegraphics[width=0.2\textwidth, valign=m]{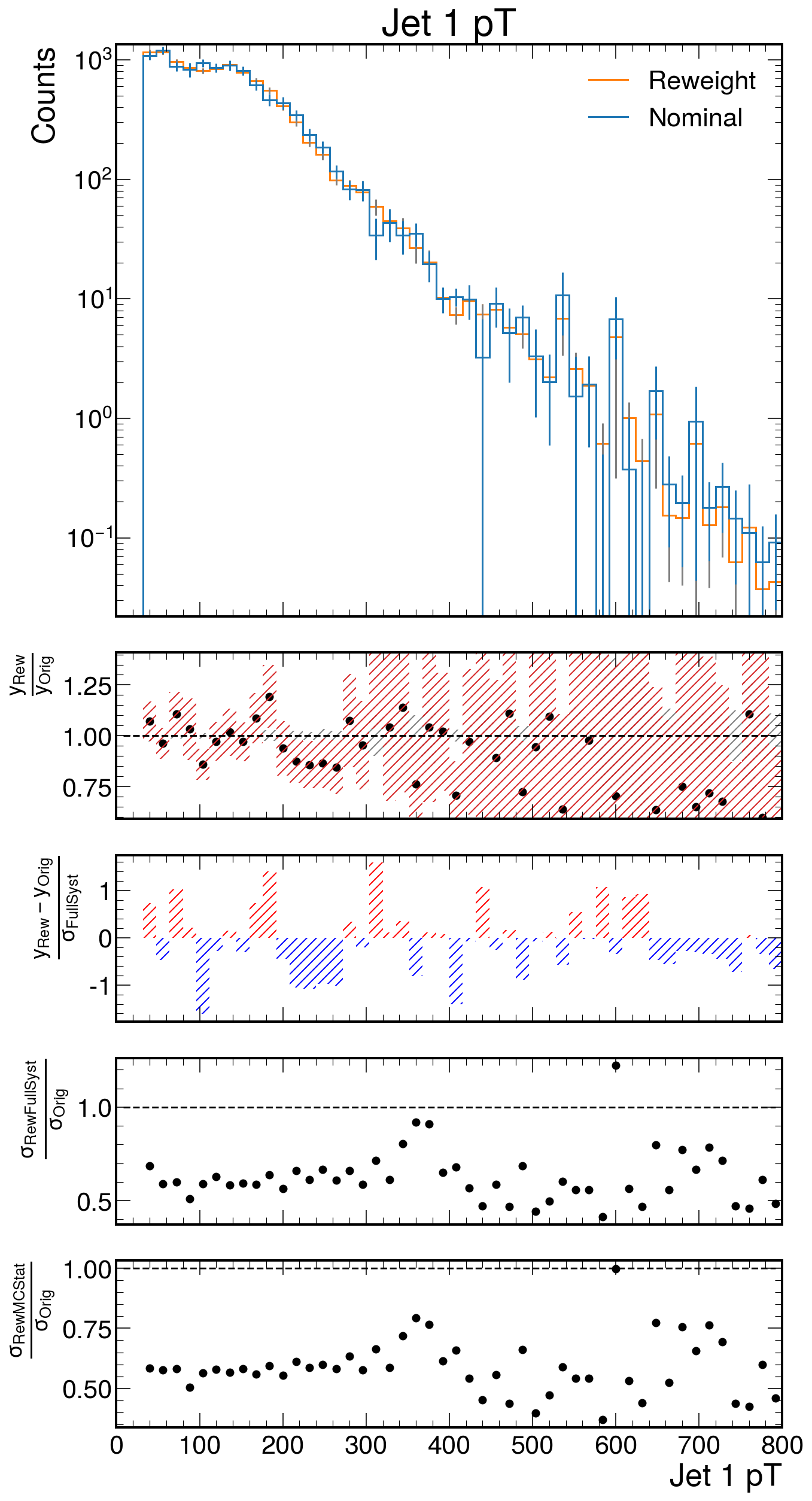}
    \includegraphics[width=0.2\textwidth, valign=m]{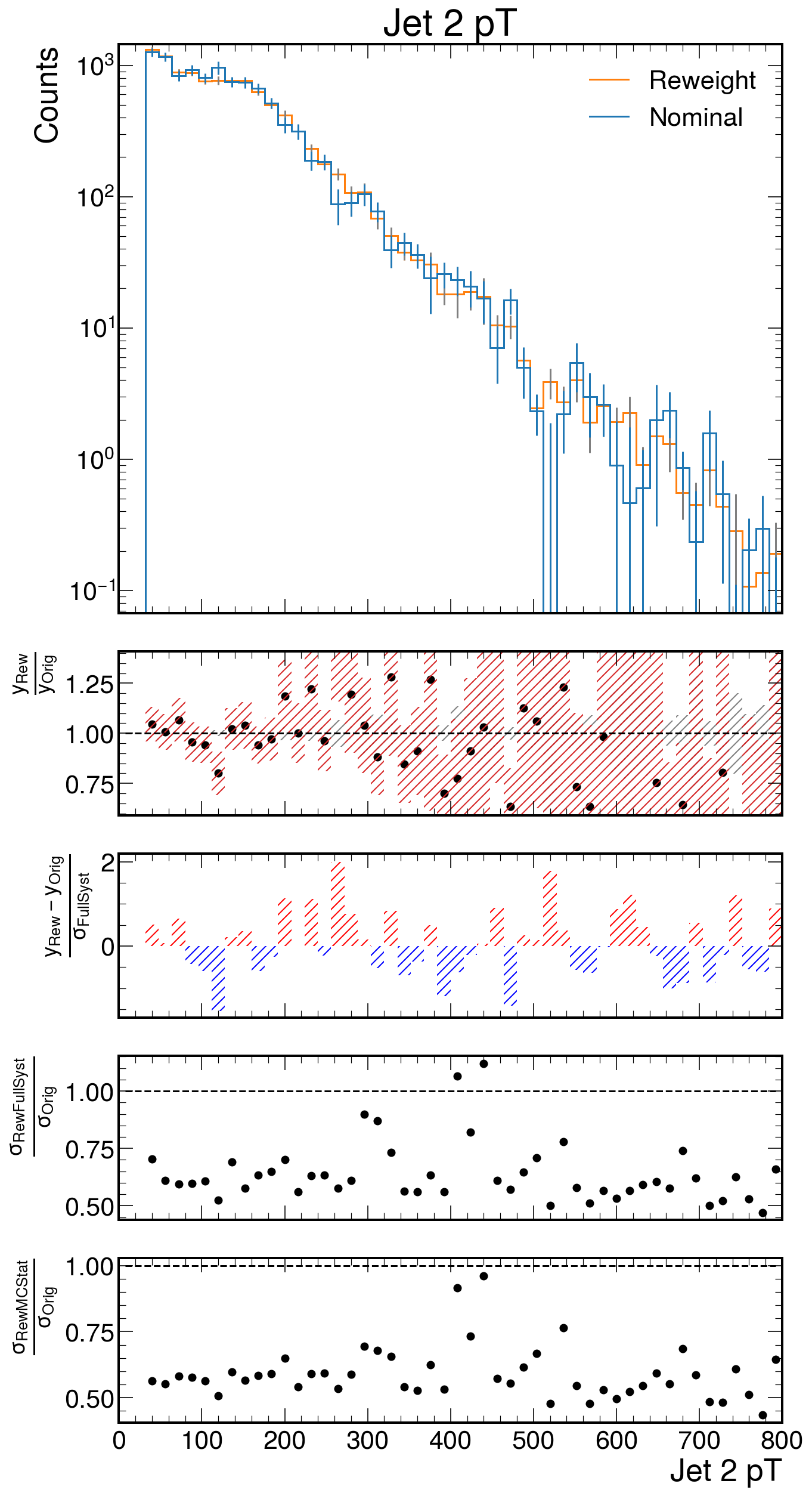}
    \caption{Plots of the reconstruction-level closure of the reweighting for the leading btag, dijet mass, and the pT of the leading and subleading jets.} 
    \label{fig:reco_validation}
\end{figure}

Finally, a few additional validation plots of various variables used to train the \dnnsvb are shown in Figure~\ref{fig:reco_validation}. All of these show good closure and a significantly reduced uncertainty.

\section{\label{sec:conclusion}Conclusion}

This paper discussed the origin of negative weights in NLO Monte Carlo samples and highlighted the problems the negative weights cause from an experimentalist’s perspective. A practical solution through the construction of a reweighting function $g(\vec{x})$ and methods to assign uncertainties to the function event-by-event and at the final observable level have been developed and shown. The PCA-based final observable level uncertainty is both more descriptive through its correlation informed shapes and faster computationally by only calculating the covariance on the histograms than the per event uncertainties.

The reweighting method was shown to reduce bin-by-bin uncertainties significantly both with and without uncertainties on the reweighting function. When the exact reweighting is known, this reduction depends only on the fraction of positive weights. This was shown mathematically, and it was demonstrated through an MC sampling performed for a double slit experiment. When the reweighting function is not known \textit{a priori}, it was shown in theory and in practice that a \dnnre can learn it by modeling the probability of an event being positive. Using public MC from ATLAS, the trained model was shown to accurately reflect the probability of events being positive. When accounting for the uncertainty using the methods proposed in the text, the added uncertainties were low enough that the overall precision in the simulation estimate were significantly improved with the reweighting in the most signal-rich bins of the \dnnsvb observable.

The methods discussed here are applicable to any MC sample with negative weights to provide a substantial improvement in their statistical precision. The PCA-based uncertainty prescription allows the proper handling of the reweighting in fits, allowing the use of the method in physics analyses. This will improve the reach of analyses and help make the most of High Luminosity LHC data when it arrives.

\begin{acknowledgments}

This work was supported in part by U.S. Department of Energy Grant No. DE-SC0010072.
We thank Sarah Eno and Harrison Prosper for their comments on the manuscript. We acknowledge the work of the ATLAS Collaboration to record or simulate, reconstruct, and distribute the Open Data used in this paper, and to develop and support the software with which it was analyzed.
\end{acknowledgments}

\section*{Data Availability}
The data that support the findings of this article are openly available \cite{ATLAS:electroweakBoson, ATLAS:higgs}.

\bibliography{main}

\end{document}